%% file: main.tex
\documentclass[11pt]{article}
\usepackage[margin=1in]{geometry}
\usepackage{natbib}
\usepackage{amsfonts}
\usepackage{amsmath}
\usepackage{amsthm}
\usepackage{amssymb}
\usepackage{mathtools}
\usepackage{hyperref}
\usepackage{nameref}
\usepackage{multirow}
\usepackage{adjustbox}
\usepackage{xcolor}
\usepackage{graphicx}
\usepackage{caption}
\usepackage{subcaption}

\newcommand{\precisionmatrix}{\mathbf{\Sigma}^{-1}}
\newcommand{\mathcalN}{\mathcal{N}}

\newtheorem{theorem}{Theorem}
\newtheorem{corollary}{Corollary}[theorem]
\newtheorem{lemma}{Lemma}

\input{mathcommand}

\title{Contraction of a quasi-Bayesian model with shrinkage priors in precision matrix estimation}
\author{Ruoyang Zhang\thanks{Email: njiandan@ufl.edu}\\ 
Statistics Department, University of Florida 
\and 
 Yisha Yao\\ Department of Statistics, Rutgers University
\and Malay Ghosh\\ 
Statistics Department, University of Florida}

\begin{document}
\maketitle

\begin{abstract}
    Currently several Bayesian approaches are available to estimate large sparse precision matrices, including Bayesian graphical Lasso \citep{wang2012bayesian}, Bayesian structure learning \citep{banerjee2015bayesian}, and graphical horseshoe \citep{li2019graphical}. Although these methods have exhibited nice empirical performances, in general they are computationally expensive. Moreover, we have limited knowledge about the theoretical properties, \textit{e.g.}, posterior contraction rate, of graphical Bayesian Lasso and graphical horseshoe. In this paper, we propose a new method that integrates some commonly used continuous shrinkage priors into a quasi-Bayesian framework featured by a pseudo-likelihood. Under mild conditions, we establish an optimal posterior contraction rate for the proposed method.  
    Compared to existing approaches, our method has two main advantages. First, 
    our method is computationally more efficient while achieving similar error rate; second, our framework is more amenable to theoretical analysis.
    Extensive simulation experiments and the analysis on a real data set are supportive of our theoretical results.
\end{abstract}

{\bf Keywords:} posterior consistency, shrinkage priors, pseudo-likelihood.

\newpage
\section{Introduction}
Precision matrices are critical in a wide range of disciplines, including social networks, biomedical sciences, and economics. The zero-nonzero structure of the precision matrix $\Omega=\precisionmatrix\in \R^{p\times p}$ corresponds to the edge pattern of a Gaussian graphical model where the nodes are distributed as $\mathcalN(\textbf{0}, \mathbf{\Sigma})$ \citep{lauritzen1996graphical}. 
An element $\Omega_{ij}$ of $\Omega$ represents the conditional covariance between node $i$ and $j$ given the remaining nodes. Thus $\Omega_{ij}=0$ implies conditional independence of the two nodes given the rest. 
Bearing these significant implications, the problem of precision matrix estimation has attracted enormous attention in the past few decades, which can be described as follows. Suppose $\bx$ is a $p$-dimensional multivariate normal random vector, $\bx=(x_1, x_2, \cdots, x_p) \sim \mathcalN(\mu, \Omega^{-1})$. Given a design matrix $\bX\in \R^{n\times p}$ whose rows are \textit{i.i.d.} copies of $\bx$, one aims to obtain an estimator of $\Omega$. 

In contemporary real-world applications, researchers frequently confront cases where the dimension $p$ of the precision matrix is comparable or even larger than the sample size $n$ and the corresponding precision matrix is sparse. Naturally, in such scenarios, one would like to exploit the sparsity of the underlying precision matrix so as to uncover the edge pattern of the graph and to measure the conditional dependence among the nodes. Yet seeking a sparse and accurate estimator can be rather difficult since one needs to address entry selection besides estimation. 

The issue of entry selection in precision matrices was raised in \citep{dempster1972covariance} which he referred to as ``covariance selection". Conventional approaches for covariance selection builds upon discrete optimization, including stepwise forward and backward search, as described in \citep{whittaker1990graphical,lauritzen1996graphical, edwards2012introduction}. Entry estimation is then based on the selected model \citep{whittaker1990graphical,lauritzen1996graphical, edwards2012introduction}.  
However, these approaches become computationally infeasible even when the dimension $p$ is moderately large. Besides, the discrete procedures may lead to selection instability \citep{breiman1996heuristics}. There have been numerous methods emerging recently that alleviate the instability of the discrete selection procedure and apply to very large graphs. In the following paragraphs we shall selectively review the existing literature for precision matrix estimation. 

First we briefly go through several frequentist approaches that emerged recently. They generally fall under four categories, Cholesky decomposition based method, penalized likelihood estimation, regularized regression, and reduced tuning estimation. \cite{huang2006covariance} proposed to reparameterize the precision matrix or covariance matrix via Cholesky decomposition, then estimate the Cholesky factor, and finally estimate the precision matrix through the Cholesky representation. 
Penalized likelihood estimators are proposed in \cite{huang2006covariance}, \cite{yuan2007model}, \cite{banerjee2008model}, \cite{friedman2008sparse}, \cite{zhang2014sparse}, among others. Due to the nontrivial constraints, \textit{e.g.}, positive definiteness, penalized likelihood methods are relatively computationally expensive. Efficient algorithms are essential for their successful implementation. The existing algorithms to maximize the penalized log likelihood function include the maxdet algorithm \citep{yuan2007model}, block coordinate descent \citep{banerjee2008model}, Nesterov's first order method \citep{banerjee2008model}, and majorization-minimization algorithm \citep{lange2000optimization, friedman2008sparse}. 
The theoretical properties, \textit{e.g.}, convergence rates, of these penalized likelihood estimators are detailed in \cite{rothman2008sparse, lam2009sparsistency, ravikumar2011high}.
Regularized-regression type methods are based on the fact that when regressing one node $x_i$ against the remaining nodes $\{x_j\}_{j\neq i}$, the theoretical regression coefficients are equal to $\{-\Omega_{ii}^{-1}\Omega_{ij}\}_{j\neq i}$ \citep{fan2016overview}. Naturally Lasso \citep{tibshirani1996regression}, Dantzig selector \citep{candes2007dantzig}, and sclaed Lasso \citep{sun2012scaled} have been employed for columnwise precision matrix estimation/selection, respectively in \cite{meinshausen2006high}, \cite{yuan2010high}, and \cite{sun2013sparse}.
Most of the above methods rely on tuning parameters that are functions of the pre-perceived parameters. 
While reduced tuning estimation, such as TIGER (Tuning-Insensitive Graph Estimation and Regression) \citep{liu2017tiger} and EPIC (Estimating Precision matrIx with Calibration) \citep{zhao2014calibrated}, is asymptotically tuning-free or requires very few efforts on tuning parameters. 

We then discuss some Bayesian approaches which are the primary focus of this paper. In addition to estimation, Bayesian model selection procedures can produce posterior distributions that quantify estimation and selection uncertainty. To promote sparsity, Bayesian approaches usually impose sparsity-inducing priors like spike-and-slab priors, which put a point mass at zero and continuous shrinkage priors including the double exponential prior (Bayesian Lasso) \citep{park2008bayesian, hans2009bayesian}, horseshoe prior \citep{carvalho2010horseshoe}, normal-gamma prior \citep{brown2010inference}, double-Pareto prior \citep{armagan2013generalized}, Dirichlet-Laplace prior \citep{bhattacharya2015dirichlet}, horseshoe+ prior \citep{bhadra2017horseshoe+}, and the continuous spike-and-slab prior \citep{rovckova2018bayesian}. Compared to the point mass prior, these continuous shrinkage priors ease the computation to a large extent. Moreover, they also enjoy reasonable concentration properties \citep{song2017nearly, wei2020contraction}.
The asymptotic normality of posteriors is established in \cite{ghosal1997normal, ghosal1999asymptotic, ghosal2000asymptotic} under the condition that $p$ grows slower than $n$, while the general theory of posterior contraction was established in \cite{ghosal2000convergence}, which provides a broadly applicable technique for more specific settings. In \cite{jiang2007bayesian}, it is shown that the posterior convergence rate of some special priors in terms of Hellinger distance can be close to $n^{-1/2}$.  \cite{castillo2012needles} studied the variable selection consistency and posterior contraction in sparse normal mean models with certain point-mass priors. \cite{castillo2015bayesian} established the optimal posterior contraction rate and selection consistency in high-dimensional linear regression models for priors that are mixtures of point masses at zero and continuous distributions. A more recent paper proved the optimal posterior contraction rate for the empirical Bayes method in high-dimensional linear regression models \citep{belitser2020empirical}. Compared to point-mass priors, it is more challenging to study the posterior contraction properties under continuous shrinkage priors. Luckily there have been several milestone papers addressing this difficult problem. \cite{armagan2013posterior} showed that the posteriors concentrate around the true parameters when $p$ grows sufficiently slow (slower than $n$) for shrinkage priors. \cite{van2014horseshoe} and \cite{bhattacharya2015dirichlet} gave optimal posterior contraction rates for the horseshoe prior and Dirichlet-Laplace prior, respectively. \cite{song2017nearly} showed that a wide class of continuous shrinkage priors would attain a posterior contraction rate similar to that of the spike-and-slab prior in high-dimensional linear regression models. There are also several papers that extend the aforementioned results to generalized linear models \citep{atchade2017contraction, wei2020contraction}, or high-dimensional nonparametric additive models \citep{yang2015minimax, shen2016adaptive,  belitser2020empirical}.

Despite such huge amount of literature on Bayesian approaches in linear models, limited work exits on exploring the theoretical properties of shrinkage priors for sparse precision matrix estimation. 
The early work by \citep{carvalho2009objective} addresses model selection in Bayesian Gaussian graphical models by combining a multiplicity-correction prior and fractional Bayes factors. 
\cite{banerjee2014posterior} imposed a conjugate graphical Wishart prior on $\Omega$ and established a contraction rate $k^{5/2}\sqrt{\log p/n}$ when the true precision matrix has $k$-banded structure, while \cite{xiang2015high} extended the results to general decomposable graphs. A rate of $k^{3/2}\sqrt{\log p/n}$ is established in \citep{lee2021estimating} for a special class of banded precision matrices using k-banded Cholesky priors. For arbitrary sparsity structures, \cite{banerjee2015bayesian} used a prior that puts a mixture of a point mass at zero and certain absolutely continuous distribution on off-diagonal elements and established a contraction rate of $\sqrt{(p+s^*)\log p/n}$ in terms of Frobenius norm, where $s^*$ is the number of nonzero off-diagonal elements in the true precision matrix.
Meanwhile, some Bayesian methods with continuous shrinkage priors like Bayesian graphical Lasso \citep{wang2012bayesian} and graphical horseshoe \citep{li2019graphical} are shown to perform well empirically, but theoretical properties are limited. More descriptions of these two methods will be given in Section 2. 

For precision matrix estimation, the Bayesian methods mentioned above are based on the multivariate Gaussian likelihood. 
Two recent papers by \cite{atchade2017contraction, atchade2019quasi} came up with a quasi-Bayesian scheme where the spike-and-slab prior is combined with a pseudo-likelihood, and the resulting quasi-posterior distribution has nice contraction properties. 
Inspired by this pseudo-likelihood function in \cite{atchade2017contraction, atchade2019quasi}, 
we propose to estimate large sparse precision matrices by integrating continuous shrinkage priors into a quasi-Bayesian scheme. 
Although we use a similar quasi-Bayesian framework as in \cite{atchade2019quasi}, there is significant difference. \cite{atchade2019quasi} studied the spike-and-slab prior within the quasi-Bayesian scheme, while we explore the theoretical properties of some common continuous shrinkage priors. 
An optimal posterior contraction rate is established for our method by extending some techniques in \cite{song2017nearly, wei2020contraction, song2020bayesian}.
To the best of our knowledge, this is the first effort to study the posterior contraction properties of a general class of continuous shrinkage priors in a quasi-Bayesian framework for precision matrix estimation.

The rest of this paper is organized as follows. Section 2 introduces some of the most relevant works and outlines our quasi-Bayesian framework. Section 3 states some assumptions and provides theoretical analysis on the posterior contraction rate. In Sections 4 and 5, we illustrate the proposed method through extensive simulation studies and analyzing a real data set. Finally in Section 6, we conclude with some discussions and possible future research topics. Proofs are relegated to the Appendix. 

\section{Methodology}
We are interested in estimating the precision matrix $\Omega\in\mathbb{R}^{p\times p}$ from the sample data $\bY\in\Rnp$, where $n$ denotes the sample size, and $p$ is the dimension of precision matrix or equivalently the number of nodes in the corresponding graph. The rows of $\bY$ are assumed to be i.i.d. $\mathcalN(\bm{0},\Omega^{-1})$ distributied. The likelihood function based on the data can be written as
\begin{equation}\label{eq:true_llh}
L(\bY|\Omega) =  (2\pi)^{-\frac{np}{2}}\det(\Omega)^{\frac{n}{2}}\exp \left \{-\frac{1}{2}\tr(\bY \Omega \bY^T) \right \}.
\end{equation}
And we will discuss this problem in a quasi-Bayesian framework, which is motivated by the following three papers.

\subsection{Related Work on Bayesian Precision Matrix Estimation}
The first is the Bayesian Graphical Lasso (BGL) presented in \cite{wang2012bayesian}. BGL puts on $\Omega$ a prior as follows:
    \[ \pi(\Omega|\lambda) \propto \prod_{i<j}\DE(\omega_{ij}|\lambda) \prod_{i=1}^p \EXP(\omega_{ii}| \lambda/2) 1_{\Omega \in \setS_p},\]
    where $\DE(\cdot|\lambda)$ and $\EXP(\cdot|\lambda)$ represent the double exponential density $p(x)=\lambda/2\exp(-\lambda |x|)$ and the exponential density $p(x)=\lambda\exp(-\lambda x)1_{x>0}$, respectively, and $\setS_p$ is the space of $p\times p$ positive definite real matrices. 

More recently, \cite{li2019graphical} came up with the Graphical Horseshoe (GHS) estimator. GHS puts the following prior on off-diagonal elements of $\Omega$:
    \[
    \pi(\Omega|\tau) \propto \prod_{i<j} \mathcalN(\omega_{ij}|\lambda_{ij}^2, \tau^2) \prod_{i<j}\halfCauchy(\lambda_{ij}|0,1) 1_{\Omega \in \setS_p},
    \]
where $\halfCauchy(\cdot|0,1)$ denotes the half-Cauchy distribution with density $p(x)=2/\pi (1+x^2)^{-1}$.
Both BGL and GHS do not restrict the sparse patterns, such as banding \citep{banerjee2014posterior}. 
Yet GHS is claimed to have better asymptotic properties compared to BGL. Unlike the BGL, when $n\succ p$ the GHS is close to an unbiased estimator for a nonzero parameter with high probability.
The posterior samples of BGL and GHS are obtained by Gibbs sampling whose computation complexity is $O(p^3)$ due to inverting a $p\times p$ matrix within each step. So BGL and GHS are generally not applicable to very large precision matrix estimation. In addition, no posterior consistency result has yet been established for either method.

\cite{atchade2019quasi} proposed a quasi-Bayesian method that can be applied to large Gaussian graphical models. It is based on a pseudo-likelihood instead of the full likelihood function \eqref{eq:true_llh}.
For a $p$ dimensional random vector $\bx=(X_1,\cdots,X_p)'$, let $\bx_{-i}$ be the random vector leaving out one element $X_i$, \textit{i.e.}, $\bx_{-i}=(X_1,\cdots, X_{i-1},X_{i+1},\cdots, X_p)'$. If $\bx \sim \mathcalN(\bm{0},\Omega^{-1})$, then the conditional distribution of $X_i$ given $\bx_{-i}$ is
\begin{equation}\label{eq:conditional_org}
    X_i | \bx_{-i}=(x_1, \cdots, x_{i-1}, x_{i+1}, \cdots, x_p) \sim \mathcalN\left( -\sum_{j\neq i} \dfrac{\omega_{ji}}{\omega_{ii}}x_j, \dfrac{1}{\omega_{ii}} \right),  \quad j\neq i,
\end{equation}
where $\omega_{ij}$ is the $ij$-th element of $\Omega$. 
Taking advantage of the conditional distribution \eqref{eq:conditional_org}, \cite{atchade2019quasi} came up with the following pseudo-likelihood function
\begin{equation}\label{eq:pseudo-llh}
    q(\bY|\Omega) = \prod_{j=1}^p q_j(\bY_{\cdot,j}|\bY_{\cdot, -j})
\end{equation}
where 
\[
q_j(\bY_{\cdot, j}|\bY_{\cdot, -j}) = \left(\dfrac{\omega_{jj}}{2\pi} \right)^{n/2}\exp\left(-\dfrac{\omega_{jj}}{2} \norm{\bY_{\cdot, j} + \sum_{k\neq j} \dfrac{\omega_{kj}}{\omega_{jj}} \bY_{\cdot, k}}^2  \right),
\] 
$\bY_{\cdot, j}$ denotes the $j$th column of the data matrix $\bY$, and $\bY_{\cdot, -j}$ is the matrix obtained by removing the $j$th column of $\bY$.
A spike-and-slab prior is imposed on $\Omega$. For each $j$, $\omega_{ij}$ is i.i.d. distributed as 
\begin{equation}\label{eq:spike_prior}
    \pi(\omega_{ij})=q\cdot Dirac(0) + (1-q)\cdot Laplace(\rho_j\omega_{jj}), \; i\neq j,
\end{equation}
where $Dirac(0)$ is the Dirac measure on $\R$ with all mass concentrated on zero, $0<q<1$, and $\rho_i$ is a hyper-parameter. The diagonal elements $\omega_{ii}$ are assumed to be known.
With the pseudo-likelihood \eqref{eq:pseudo-llh} and prior \eqref{eq:spike_prior}, the quasi-posterior distribution is obtained as
\[
    \check{\Pi}(\Omega|\bY)=\prod_{j=1}^p \check{\Pi}(\Omega_{-j,j} | \bY)
    \propto \prod_{j=1}^p q_j(\bY_{\cdot, j}|\bY_{\cdot, -j}) \pi(\Omega_{-j,j}),
\]
where $\Omega_{-j,j}$ denotes the $j$th column of $\Omega$ without $j$th row. 
\cite{atchade2019quasi} shows that the above quasi-posterior distribution contracts at a rate of $\bigO(d_*\sqrt{\log p /n})$ in terms of the spectral norm, where $d_*$ is the maximum degree of the corresponding graph. This rate matches the optimal convergence rate of the frequentist approaches for neighborhood selection \citep{sun2013sparse}.


\subsection{A Quasi-Bayesian Model with Shrinkage Priors}
In addition to the spike-and-slab priors, some continuous shrinkage priors have exhibited both nice empirical performances \citep{carvalho2010horseshoe, polson2010shrink} and nearly optimal posterior contraction rates \citep{song2017nearly} in mean estimation and linear regression problems.
Motivated by the good theoretical properties and computational feasibility of these continuous shrinkage priors, we propose the following Quasi-Bayesian model with Shrinkage Priors (QBSP). Similar to \cite{atchade2019quasi}, we assume the diagonal elements $\omega_{ii}$ of $\Omega$ are known. Without loss of generality, we assume the diagonal $\omega_{ii}=1$ for all $i=1,\ldots,p$.

QBSP uses the pseudo-likelihood in \eqref{eq:pseudo-llh} with the following prior on $\Omega$: 
\begin{equation}\label{eq:QBSP_prior}
    \omega_{ij}\sim\pialpha(\omega_{ij}), \; j\neq i \quad \text{independently},
\end{equation}
where $\pialpha(\cdot)$ satisfies
\begin{subequations}\label{cond:prior}
\begin{align}
     1-\int_{-a_n}^{a_n}\pialpha(x) dx & \leq p^{-(1+u)} \; \text{for some constant } u>0, \label{cond:priorconcentration}\\
     \inf_{x\in[-E_n, E_n]} \pialpha(x)  & \geq  p^{-c} \; \text{for some constant } c>1.
    \label{cond:fattail}
\end{align}
\end{subequations}
Here, $\alpha$ is the hyper-parameter. In \eqref{cond:priorconcentration}, $a_n\leq \epsilon_n/p$, where $\epsilon_n$ is the contraction rate that varies in different problems and will be determined later. Note that $\epsilon_n$ satisfies $\epsilon_n \to 0$ and $n\epsilon_n^2 \to \infty$. The $E_n$ in \eqref{cond:fattail} is an upper bound for the true signal strength $|\omega^*_{ij}|$. Specific formula of these quantities will be given in different scenarios as discussed below in Section 3.

The two conditions \eqref{cond:priorconcentration} and 
\eqref{cond:fattail} were originally given in \cite{song2017nearly} to tackle the linear regression problem under Bayesian settings. \eqref{cond:priorconcentration} essentially states that the majority mass of the prior distribution lies within a tiny interval $[-a_n, a_n]$. This high concentration of the prior density around zero mimics the spike-and-slab prior in promoting model sparsity. While \eqref{cond:fattail} implies that the prior density around the true value should be at least $p^{-c}$, \textit{i.e.}, the prior is ``thick" enough around the true parameter value. Since the maximum nonzero entry of $\Omega$ is usually bounded away from zero, this condition suggests that the tail of prior distribution need to be heavy enough.

For the moment, the conditions given in \eqref{cond:prior} might seem abstract and not easy to interpret. We will give some specific examples in Section 3, and show that many commonly used shrinkage priors actually satisfy the conditions in \eqref{cond:prior}. One may notice that the prior is not necessarily symmetric or positive definite. As a consequence, the corresponding posterior samples are generally not symmetric or positive definite, while they should be. Luckily, the posterior consistency established in Theorem \ref{thm:ggm} actually implies the  positive definiteness of posteriors with high probability. That leaves the symmetricity issue to be solved, which we shall discuss at the end of Section 3.

\section{Contraction Results}
In this section, we establish the posterior consistency of the proposed quasi-Bayesian model with shrinkage priors. First, let us introduce some notations. For two positive sequences $a_n$ and $b_n$, $a_n \prec b_n$ is equivalent to $b_n \succ a_n$, which means $a_n/b_n\to 0$ as $n\to\infty$; $a_n \simeq b_n$ means that $c_1< a_n/b_n < c_2$ for all large $n$ and some constants $c_1, \; c_2 >0$. $a_n = \bigO(b_n)$ denotes $a_n/b_n \leq C$ for all large $n$ and a constant $C>0$ . For a vector $\bv\in\Rp$ and a real number $r>0$, $\norm{\bv}_r = \left( \sum_{j=1}^p|v_j|^r \right)^{1/r}$ and $\norm{\bv}_0=\sum_{j=1}^p I(v_j\neq 0)$. For a matrix $\bm{A}\in\R^{p\times p}$, $\normF{\bm{A}}=\sqrt{trace(\bm{A}^T\bm{A})}$. 

To utilize the conditional distribution in \eqref{eq:conditional_org} and the pseudo-likelihood scheme in \eqref{eq:pseudo-llh}, we first consider the following linear regression model:
\begin{equation}\label{eq:linearreg}
    \bZ = \bX\bm{\theta} + \bm{\varepsilon}, \; \bm{\varepsilon}\sim N(0,I_n).
\end{equation}
Model \eqref{eq:linearreg} is consistent with the conditional distribution in \eqref{eq:conditional_org} given that $\omega_{ii}=1$ for all $i$.
Here $\bX\in\mathbb{R}^{n\times (p-1)}$ is a random design matrix with its rows i.i.d $\mathcalN(0,\Sigma_{X})$ distributed, and $\bm{\theta}\in\mathbb{R}^{p-1}$ is the regression coefficients to be estimated. Rigorously, the column dimension of $\bX$ and the dimension of $\bm{\theta}$ should be $p-1$. However, it does not affect any of our results to use $p$ instead of $p-1$.
So we will just use $p$ in subsequent discussion for simplicity.

Before elaborating Theorem \ref{thm:linearreg}, we give the following assumptions about dimension $p$, sample size $n$ and true coefficient $\thetastar$ based on which Theorem \ref{thm:linearreg} is established.
\begin{itemize}
    \item[] $C_1(1)$: The dimension $p$ grows to $\infty$ as $n\to\infty$ and $\log p \simeq \log n$;
    \item[] $C_1(2)$: $\sstar\log p \prec n$, where $\sstar$ is the size of true model, \textit{i.e.}, $\sstar=\normzero{\thetastar}$;
    \item[] $C_1(3)$: $\max\{|\theta^*_{i}|\} \leq \gamma E_n$ for fixed constant $\gamma\in(0,1)$, and $E_n$ is non-decreasing with $n$.
\end{itemize}
The above conditions seem similar to those given in \cite{song2017nearly}. However, there are several important differences between their work and ours.
For example, \cite{song2017nearly} is working with uniformly bounded design matrices while we are dealing with multivariate Gaussian designs that are not bounded; and they are addressing the linear regression problem while our ultimate goal is to estimate large precision matrix.

\begin{theorem} \label{thm:linearreg} Consider the linear regression model \eqref{eq:linearreg}. Suppose $n$, $p$ and the true $\thetastar$ satisfy conditions $C_1$. The prior is given by $\pialpha(\bm{\theta})=\prod_{i=1}^p\pialpha(\theta_i)$, where $\pialpha(\theta_i)$ satisfies (\ref{cond:prior}). Then the following posterior consistency results hold with probability at least $1-\exp(-cn)$,
\begin{subequations}\label{pstcon:linear}
\begin{align}
        \Estar[\Pi(\sum_{i=1}^p I(\theta_i > a_n) > Ls^*)| \bZ] & \leq 7/p^2 \label{pstcon:linear_selection}\\
    \Estar[\Pi(\normtwo{\bm{\theta}-\thetastar}\geq M\epsilon_n | \bZ] & \leq 7/p^2 \label{pstcon:linear_est}
\end{align}
\end{subequations}
where $\epsilon_n=\sqrt{\sstar\log p/n}$, $c$, $L$ and $M$ are positive absolute numerical constants. 
\end{theorem}

The proof of Theorem \ref{thm:linearreg} is given in the Appendix. 
Since the shrinkage prior is continuous, the posterior samples are generally not exactly zeros. However, the prior distribution is highly concentrated within a tiny interval $[-a_n, a_n]$, which closely mimics the point mass at zero of the spike-and-slab distribution. From this point of view, $a_n$ somehow distinguishes zero and nonzero coefficients in the prior. Therefore, it is natural to consider the posterior variable selection rule $\tilde{S}(\bm{\theta}|\bZ)=\{i: |\theta_i| > a_n | \bZ \}$. Under this selection rule, \eqref{pstcon:linear_selection} implies that the posterior model size is at most $L\cdot s^* = \bigO(s^*)$ with high probability, where $s^*$ is the true model size. 
\eqref{pstcon:linear_est} shows that the $\ell_2$ contraction rate of the posterior is $\bigO(\sqrt{\sstar\log p/n})$, which matches the nearly optimal rate in existing literature \citep{song2017nearly}. 

Equipped with Theorem \ref{thm:linearreg}, we are ready to establish the posterior consistency of the quasi-Bayesian method with shrinkage priors satisfying \eqref{eq:QBSP_prior} and \eqref{cond:prior} on precision matrix estimation and variable selection. Similarly, we first give the following regularity conditions on $p$, $n$ and the true parameter $\Omegastar=(\omega^*_{ij})$.
\begin{itemize}
    \item[] $C_2(1)$: The dimension $p$ grows to $\infty$ as $n\to\infty$, $\log p \simeq \log n$ and $p\log p \prec n$ ;
    \item[] $C_2(2)$: $S^* \log p \prec n$, where $S^*$ is the number of nonzero off-diagonal elements in the true precision matrix $\Omegastar$;
    \item[] $C_2(3)$: $0<k \leq \lambda_{\min}(\Omegastar) \leq \lambda_{\max}(\Omegastar) \leq 1/k$, where $\lambda_{\min}$ and $\lambda_{\max}$ refer to the minimum and maximum eigenvalue of $\Omegastar$;
    \item[] $C_2(4)$: $\max\{|\omega^*_{ij}|\} \leq \gamma E_n$ for a fixed constant $\gamma\in(0,1)$, and $E_n$ is non-decreasing with $n$.
\end{itemize}

\begin{theorem} \label{thm:ggm}
Suppose the data $\bY\in\Rnp$ has i.i.d rows drawn from the multivariate Gaussian distribution $\mathcalN(0, \Omega^{-1})$, and the pseudo-likielihood $\eqref{eq:pseudo-llh}$ is used. Further assume $n$, $p$ and the true precision matrix $\Omegastar$ satisfy conditions $C_2$, and the prior $\pi(\Omega)$ follows \eqref{eq:QBSP_prior}. Then the posterior consistency of the quasi-posterior $\tilPi(\cdot|\bY)$ holds as follows,
\begin{subequations}\label{pstcon:ggm}
\begin{align}
\EstarOmega[\tilPi(\sum_{i=1, i\neq j}^p I(\omega_{ij}>a_n) > L \sstar_j \text{ for some } j | \bY)] & \leq \exp(-c_5 n) + 7/p \label{pstcon:ggm_selection}\\
\EstarOmega[\tilPi(\normF{\Omega-\Omegastar} > M\epsilon_n | \bY)] & \leq \exp(-c_5 n) + 7/p \label{pstcon:ggm_est}
\end{align}
\end{subequations}
where $\epsilon_n=\sqrt{(\sstar_1+\cdots+\sstar_p)\log p/n}$,  $\sstar_i$ is the number of nonzero off-diagonal elements in the $i$th column of $\Omegastar$. $L$, $M$ and $c_5$ are positive absolute constants.
\end{theorem}

Since $S^* = \sstar_1+\ldots+\sstar_p$, the contraction rate of $\tilde{\Pi}(\cdot|\bY)$ equals to $\bigO(\sqrt{S^*\log p/n})$, which is sharper than $\sqrt{(p+S^*)\log p/n}$ in \cite{banerjee2015bayesian}, especially when $S^*\prec p$. However, it comes with a price of extra information about the nonzero elements in every row/column of $\Omegastar$, or equivalently, the degree of each node in the graph. The improvement in the contraction rate can be attributed to some extra structural information about the graph, as is in \cite{atchade2019quasi}. 

However, if we do not know the number of nonzero elements in each column and only have information about $S^*$, the contraction rate will be
\begin{equation}
    \epsilon_n=\left( (\max(\sstar_1,1)+\ldots+\max(\sstar_p,1))\log p/n \right)^{1/2}.
\end{equation}
Note that $\max(p,S^*) \leq \max(\sstar_1,1)+\ldots+\max(\sstar_p,1) \leq p+S^*$, and hence $\epsilon_n \simeq \sqrt{(p+S^*)\log p/n}$. This rate is comparable with the posterior convergence rate in \cite{banerjee2015bayesian} as well as the convergence rate of some frequentist penalized likelihood approach, \textit{e.g.}, the SPICE estimator \citep{rothman2008sparse}.

Theorem \ref{thm:ggm} gives a posterior contraction rate for the pseudo-likelihood scheme \eqref{eq:pseudo-llh} combined with the priors satisfying general conditions \eqref{cond:prior}. The conditions in \eqref{cond:prior} may seem general and abstract. 
In practice, a wide range of prior distributions both satisfy these conditions and enable efficient posterior sampling. As an example, we provide a class of global-local shrinkage priors that can be represented as the scale mixtures of Gaussians:
\begin{equation}
    \label{scale_mix_gaussian}
    \theta_i \sim \mathcalN(0, \alpha \xi_i), \quad \xi_i\sim\pi(\xi_i),
\end{equation}
where $\alpha$ is the hyper-parameter controlling the global shrinkage level, and $\xi_i$ is a parameter shrinking individual $\theta_i$'s.

Many commonly used shrinkage priors, like the Student's t, the hypergeometric inverted beta \citep{polson2012half}, the generalized double Pareto \citep{armagan2013generalized}, the horseshoe+ \citep{bhadra2017horseshoe+}, and the “Three Parameter Beta Normal” (TPBN) mixture family \citep{armagan2011generalized} (which includes the Horseshoe \citep{carvalho2010horseshoe} and the normal-exponential-gamma \citep{brown2010inference} priors), are all scale mixtures of Gaussians. The mixing density $\pi(\xi)$ for these priors are listed in Table 1 of \cite{zhang2019ultra}. Theorem 3 of \cite{zhang2019ultra} shows that all the scale mixtures of Gaussians priors mentioned above satisfy \eqref{cond:prior} with hyper-parameter $\alpha=\bigO(a_n^2p^{-(1+u')/(r-1)})$ for some $u'>0$, $r>1$ and $-\log p = \bigO(\log\alpha)$. In subsequent numerical simulations and real data analysis, we will implement the Horseshoe prior to demonstrate using shrinkage priors in the quasi-Bayeisan framework.

Suppose the quasi-posterior distribution we obtained is $\tilde{\Pi}(\Omega|\bY)$. Note that the support of $\tilde{\Pi}(\Omega|\bY)$ is generally not restricted to the set of symmetric matrices. Although the contraction in Frobenius norm shown in \eqref{pstcon:ggm_est} implies that the posterior deviates from $\Omegastar$ by only a tiny amount, we will not stop here. To further produce a symmetric estimator, we adopt the following symmetrization step as in \cite{yuan2010high} and  \cite{sun2013sparse}:
\begin{equation}\label{eq:linearprog}
 \hat\Omega = \argmin_{\Omega \in \R^{p\times p}: \Omega = \Omega^T} \norm{\Omega-\Bar{\Omega}}_{\ell_1},
\end{equation}
where $\Bar{\Omega}$ denotes random samples from posterior distribution $\tilde{\Pi}(\Omega|\bY)$, and $\norm{\cdot}_{\ell_1}$ is the induced operator norm defined as 
\[
\norm{A}_{\ell_1}:=\max_{x\neq 0} \dfrac{\norm{Ax}_1}{\norm{x}_1} 
= \max\limits_j \sum\limits_{i=1}^p |a_{ij}|.
\]

There is generally no closed form solution to \eqref{eq:linearprog}, but it can be solved by linear programming \citep{yuan2010high}. We give in Corollary \ref{cor}  the contraction rate of $\hat\Omega$ in terms of the spectral norm $\norm{\cdot}_{\ell_2}$.

\begin{corollary}\label{cor}
Let $\hat\Omega$ be the estimator obtained from \eqref{eq:linearprog}. Under the same setting as in Theorem \ref{thm:ggm}, one gets
\begin{equation}\label{pstcon:ggm_l1}
\EstarOmega[\tilPi(\{\norm{\hat\Omega-\Omegastar}_{\ell_2} > M d_*\sqrt{\log p/n}\} | \bY)] \leq \exp(-c_5 n) + 14/p,
\end{equation}
where $d_* = \max(\sstar_1,\ldots,\sstar_p)$, and $M$ is a positive absolute constant.
\end{corollary}

The contraction rate $\bigO(d_*\sqrt{\log p/n})$ of $\tilde{\Pi}$ in the spectral norm is consistent with the rate in \cite{atchade2019quasi} which uses spike-and-slab prior. And it also matches rates of some frequentist methods \citep{yuan2010high, sun2013sparse}. In the settings where $p$ is much larger than $d_*$, \textit{e.g.}, $p \gg d_*^2$, the contraction rate in spectral norm will be sharper than that in Frobenius norm.

\section{Simulation Study}
In this section, we implement the quasi-Bayesian scheme with shrinkage priors proposed above with some simulated data sets. As stated in Section 3, many commonly used shrinkage priors are applicable in our scheme. Here we use the Horseshoe prior listed below and refer to the corresponding model as quasi-graphical horseshoe (quasiGHS): 
\begin{equation} \label{prior:horshoe}
    \begin{aligned}
    \omega_{ij}| \lambda_{ij},\tau & \sim N(0,\lambda_{ij}^2\tau^2), \; i\neq j \\
    \lambda_{ij} &\sim C^+(0,1) \\
    \tau &\sim C^+(0,1).
    \end{aligned} 
\end{equation}

In previous sections, diagonal element $\omega_{ii}$ are assumed to be known, which is sometimes not realistic. In the simulation study, we explored both cases where $\omega_{ii}$ is known and unknown. To differentiate the two cases, the scheme with known diagonal entries is referred to as ``quasiGHS-diag". When the diagonal entries are unknown, we estimate $\omega_{ii}$ as the inverse of mean squared error of regressing $Y_i$ against the rest covariates by Lasso:
\[
\hat{\omega}_{ii} = \big( \dfrac{1}{n-\hat{s}_{\lambda}}\normtwo{\bY_{\cdot,i}-\bY_{\cdot,-i}\hat{\beta}_\lambda}^2 \big)^{-1}.
\]
A similar method has been employed in  \cite{atchade2019quasi}.

We will compare quasiGHS and quasiGHS-diag, respectively, with four existing methods, GL, GSCAD, BGL and GHS. Among them, BGL and GHS are Bayesian methods as discussed in Section 2, while GL and GSCAD are frequentist methods described as follows.
\begin{itemize}
    \item \textit{Graphical lasso\citep{friedman2008sparse}}. Graphical lasso (GL) maximizes the penalized likelihood: 
    \[ L_\lambda(\Omega|\bY) = \log(\det(\Omega) - \tr(S\Omega/n) - \sum_{i\neq j} \lambda |\omega_{ij}| \]
    where $\Omega$ is symmetric, $S=\bY^T\bY$ and $\lambda$ is regularizing tuning parameter. 
    
    \item \textit{Graphical SCAD\citep{fan2009network}}. Graphical SCAD (GSCAD) maximizes the penalized likelihood:
    \[ L_\lambda(\Omega|\bY) = \log(\det(\Omega) - \tr(S\Omega/n) - \sum_{i\neq j} \phi_{\lambda}( |\omega_{ij}|), \]
    where $\phi_\lambda(\cdot)$ is smoothly clipped absolute deviation (SCAD) penalty \citep{fan2001variable} satisfying $\phi'_\lambda(|x|) = \lambda \left\{ 1_{|x|\leq\lambda} + \dfrac{(a\lambda-|x|)_+}{(a-1)\lambda} 1_{|x|>\lambda} \right\}$ for $a>2$ and $\lambda>0$.
    
    
\end{itemize}

\begin{figure}[ht]
    \centering
    \includegraphics[width=5.3in, height=3in]{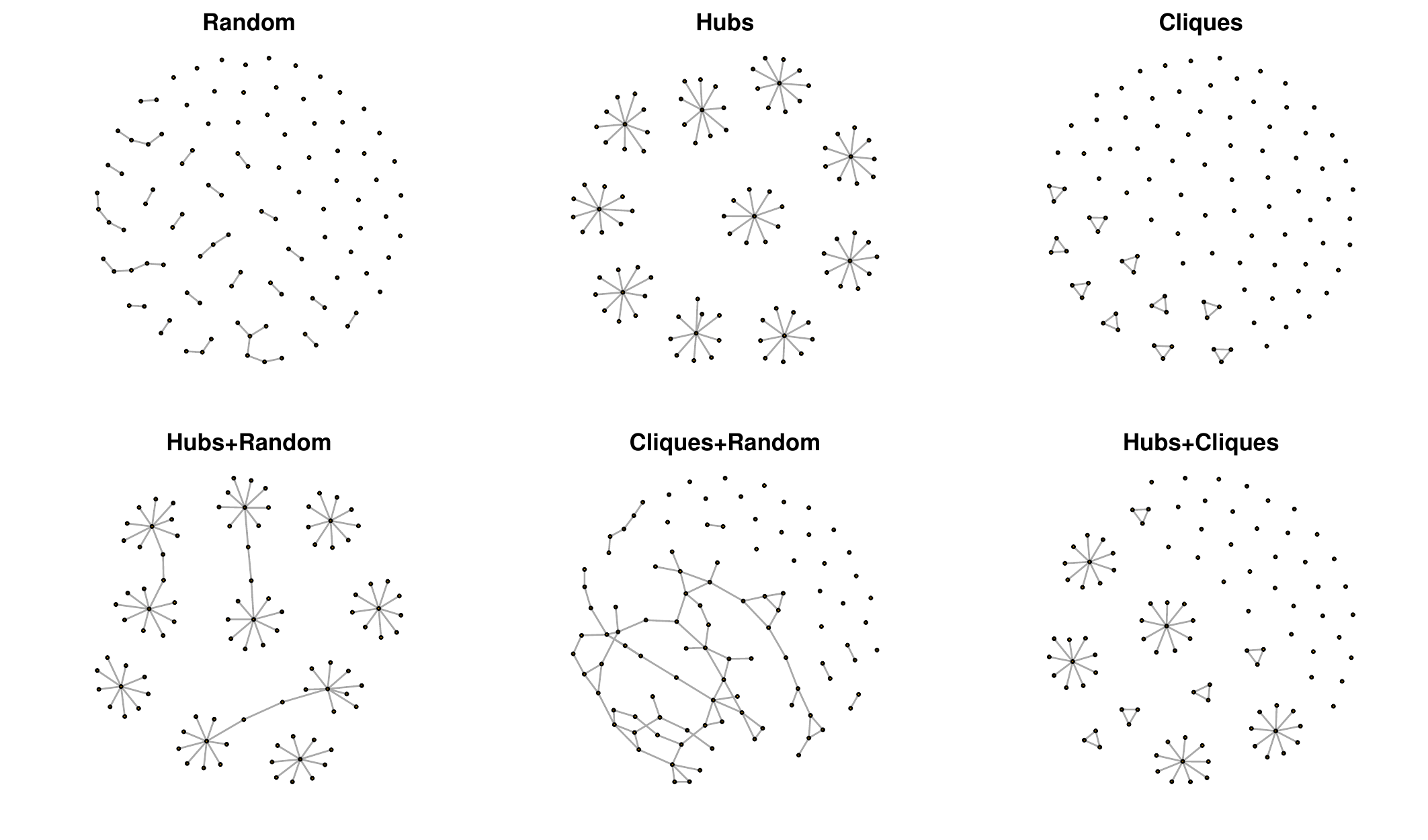}
    \caption{Six patterns of true precision matrix $\Omegastar$}
    \label{fig:6patterns}
\end{figure}

In the simulation, we set $p=100$ and $n=150$. The true diagonal elements $\omega^*_{ii}$ are set to be one, and
the true off-diagonal elements are sparse and follow one of the six patterns below. The first three patterns, \textit{Random}, \textit{Hubs} and \textit{Cliques} are similar to those in \cite{li2019graphical, friedman2010applications}. The last three patterns can be viewed as the combinations of the first three patterns. We will just call the number of nonzero elements in its upper triangular as ``the number of nonzero off-diagonal elements" since $\Omegastar$ is symmetric. Figure \ref{fig:6patterns} depicts the graphs respectively generated by the precision matrices of the six patterns. For each pattern, $50$ data sets are generated.
\begin{itemize}
    \item[--] \textit{Random.} 
    Each nonzero off-diagonal element is randomly selected with probability $1/p$ and the magnitude follows $\omega_{ij}\sim -Unif(.2, .8)$. Others are set to be zero. \\
    For $p=100$, the precision matrix we generated has $36$ off-diagonal nonzero elements.
    
    \item[--] \textit{Hubs.} The rows/columns are partitioned into disjoint groups $\{G_k\}_1^K$. Each group $G_k$ has a hub member $h_k$ such that $\omega_{i h_k}=0.25$ for $i\in G_k$ and $\omega_{ij}=0$ otherwise. \\
    For $p=100$, the $100$ variables are partitioned into $10$ groups and each group has $10$ members. There are $90$ nonzero off-diagonal elements.

    \item[--] \textit{Cliques.} The rows/columns are partitioned into disjoint groups $\{G_k\}_1^K$ with $m$ members in each group. In each group, the elements $\omega_{ij: \; i,j \in \{1,\ldots,m\}, \; i\neq j}$ are set to $-0.45$, while others are set to zero. \\ 
    For $p=100$, we consider $10$ groups and three members within each group, giving $30$ nonzero off-diagonal elements.
    
    \item[--] \textit{Hubs+Random.} In addition to the within group edges in \textit{Hubs} pattern, some random edges between groups are added in the \textit{Hubs+Random} pattern. Firstly, group pairs $k_1 \& k_2$ ($k_1,k_2 \in \{1,\ldots,K\}, \; k_1\neq k_2$) are selected with probability $1/K$, then we randomly choose $i\in G_{k_1}$, $j\in G_{k_2}$ and set $\omega_{ij}\sim -Unif(.2, .8)$.\\
    For $p=100$, there are $90$ within group nonzero elements and $3$ between group nonzero elements.
    
    \item[--] \textit{Cliques+Random.} We first partition the rows/columns into disjoint group $\{G_k\}_1^K$ with $m$ members in each group and set $\omega_{ij}=-0.3$ for $i$ and $j$ in the same group. The remaining elements are set to $0.2$ with probability $1/p$. \\
    For $p=100$, there are $87$ nonzero elements, $30$ follows \textit{Cliques} pattern and the rest $57$ are randomly selected. 
    
    \item[--] \textit{Hubs+Cliques.} The rows/columns are partitioned into disjoint groups $\{G_k\}_1^K$. Half of the groups follow \textit{Hubs} pattern with nonzero elements equal to $-0.2$ while the remaining half follow \textit{Cliques} pattern with nonzero elements equal to $0.5$.\\
    For $p=100$, there are $60$ nonzero elements, $45$ from \textit{Hubs} groups and $15$ from \textit{Cliques} groups.
\end{itemize}

We use the R package \texttt{CVglasso}\citep{rpackage_cvglasso} to implement the GL, and determine the tuning parameter by five-fold cross validation. 
For GSCAD, the R package \texttt{GGMncv}\citep{rpackage_ggmncv} is used. The tuning parameter $\lambda$ is selected by the Bayesian information criterion (BIC) criteria, and $a$ in $\phi_\lambda(\cdot)$ is set to 3.7 as recommended in \cite{fan2001variable}. We use the sampler in \cite{wang2012bayesian} to implement the BGL. The hyper-parameter $\lambda$ follows a gamma distribution with shape parameter equal to 1 and rate equal to 0.01. The sampling algorithm for GHS is available in \cite{li2019graphical} with the hyperprior $\tau \sim \halfCauchy(0,1)$.
The detailed Gibbs sampler for quasiGHS is provided in Supplementary Materials. When implementing the Bayesian methods (BGL, GHS, quasiGHS), we run the MCMC chains for 6000 iterations, and the first 1000 steps are burn-in. Convergence diagnostics of the MCMC chains are provided in the Supplementary Materials. To reduce the auto-correlation of the posterior samples, MCMC chains are thinned by 10. The R code for simulations is available on \url{https://github.com/royazhang}. And the computation is conducted on a server with Intel Xeon E5-2698 v3 CPUs at 2.30GHz without parallel computing.

To compare the estimation performance of quasiGHS and quasiGHS-diag with those of GL, GSCAD, BGL, and GHS, $\normF{\hat\Omega-\Omegastar}$ will be evaluated. For GL and GSCAD, $\hat\Omega$ is the estimate that maximizes the corresponding penalized likelihood. For BGL and GHS, the posterior mean is used as a point estimator $\hat\Omega$. For quasiGHS-diag and quasiGHS, we get the posterior means first and then apply the symmetrization step \eqref{eq:linearprog} to obtain $\hat\Omega$. 

To evaluate the variable selection performances of quasiGHS and quasiGHS-diag, we compare their true positive rates (TPR), false positive rates (FPR) and Receiver Operating Characteristic (ROC) curves with those of GL, GSCAD, BGL, and GHS. 
The TPRs (FPRs) of GL and GSCAD are calculated as the ratios of correctly selected elements (falsely selected elements) to the true nonzero elements since GL and GSCAD produce sparse estimators.
However, BGL, GHS, quasiGHS-diag and quasiGHS do not produce exact zero elements as the prior is continuous. For these Bayesian shrinkage methods, the symmetric $50\%$ posterior credible intervals are used for variable selection. If the $50\%$ posterior credible interval for $\omega_{ij, \; i\neq j}$ contains zero, the element is considered to be zero, and vice versa. In quasiGHS-diag and quasiGHS, the entry $\omega_{ij}$ will be considered nonzero if either $\omega_{ij}$ or $\omega_{ji}$ is nonzero. 
We record in Table \ref{tbl:p100n150} the means and standard deviations of TPR, FPR and $\normF{\hat\Omega-\Omegastar}$ computed over 50 data sets. The last column of Table \ref{tbl:p100n150} list the average CPU time in minutes for each method. 


\begin{table}[!ht]
\centering
\caption{Means and sds of Frobenius norm, TPR and FPR of estimate of $\Omega$ over $50$ data sets generated by $\mathcalN(0,\Omegastar^{-1})$, where $p=100$ and $n=150$. CPU time in minutes.\label{tbl:p100n150}}
\begin{tabular}{clcccc}
\hline\hline
                                    &                      &   $\normF{\hat\Omega-\Omegastar}$                        & TPR (\%)                         & FPR (\%)                        & CPU time \\ \hline
\multirow{6}{*}{Random}             & GL                   & 2.49 (0.23)                     & 97.18 (6.10)                     & 6.28 (4.24)                     & 0.08     \\
                                    & GSCAD                & 2.83 (0.16)                     & 95.83 (3.04)                     & 0.71 (0.28)                     & 0.09     \\
                                    & BGL                  & 3.64 (0.13)                     & 99.72 (1.01)                     & 19.38 (0.62)                    & 30.17    \\
                                    & GHS                  & 1.69 (0.13)                     & 92.67 (3.49)                     & 0.10 (0.05)                     & 22.49    \\
                                    & quasiGHS-diag         & 0.96 (0.09)                     & 92.56 (3.78)                     & 0.14 (0.06)                     & 7.10     \\
                                    & quasiGHS              & 1.92 (0.12)                     & 93.17 (3.23)                     & 0.19 (0.08)                     & 7.21     \\ \hline
\multirow{6}{*}{Hubs}               & GL                   & 2.27 (0.14)                     & 99.78 (0.50)                     & 7.06 (5.08)                     & 0.13     \\
                                    & GSCAD                & 1.99 (0.09)                     & 96.22 (2.90)                     & 1.18 (0.38)                     & 0.10     \\
                                    & BGL                  & 3.94 (0.14)                     & 99.84 (0.45)                     & 20.66 (0.77)                    & 29.19    \\
                                    & GHS                  & 2.16 (0.11)                     & 94.82 (2.44)                     & 0.42 (0.12)                     & 20.48    \\
                                    & quasiGHS-diag         & 2.06 (0.10)                     & 93.76 (2.36)                     & 0.45 (0.11)                     & 7.58     \\
                                    & quasiGHS              & 2.62 (0.12)                     & 94.84 (2.49)                     & 0.52 (0.14)                     & 7.76     \\ \hline
\multirow{6}{*}{Cliques}            & GL                   & 2.19 (0.26)                     & 100 (0)                          & 4.16 (3.24)                     & 0.10     \\
                                    & GSCAD                & 5.01 (0.31)                     & 100 (0)                          & 0.72 (0.23)                     & 0.10     \\
                                    & BGL                  & 3.47 (0.11)                     & 100 (0)                          & 17.65 (0.68)                    & 28.29    \\
                                    & GHS                  & 1.57 (0.17)                     & 99.87 (0.66)                     & 0.05 (0.04)                     & 20.95    \\
                                    & quasiGHS-diag         & 0.78 (0.13)                     & 99.80 (0.80)                     & 0.07 (0.04)                     & 7.63     \\
                                    & quasiGHS              & 1.80 (0.19)                     & 99.53 (1.17)                     & 0.13 (0.07)                     & 7.73     \\ \hline
\multirow{6}{*}{Hubs   + Random}    & GL                   & 2.36 (0.19)                     & 98.96 (1.78)                     & 5.62 (4.32)                     & 0.12     \\
                                    & GSCAD                & 2.05 (0.08)                     & 95.44 (2.44)                     & 1.22 (0.37)                     & 0.08     \\
                                    & BGL                  & 3.94 (0.11)                     & 99.80 (0.42)                     & 20.69 (0.69)                    & 28.98    \\
                                    & GHS                  & 2.18 (0.14)                     & 94.06 (2.48)                     & 0.48 (0.12)                     & 21.37    \\
                                    & quasiGHS-diag         & 2.07 (0.11)                     & 93.12 (2.88)                     & 0.47 (0.12)                     & 8.10     \\
                                    & quasiGHS              & 2.48 (0.19)                     & 93.46 (2.70)                     & 0.57 (0.15)                     & 7.87     \\ \hline
\multirow{6}{*}{Cliques   + Random} & GL                   & 2.49 (0.15)                     & 83.44 (9.26)                     & 5.16 (4.56)                     & 0.12     \\
                                    & GSCAD                & 2.23 (0.08)                     & 62.94 (6.60)                     & 0.38 (0.21)                     & 0.08     \\
                                    & BGL                  & 3.87 (0.13)                     & 95.26 (2.19)                     & 20.07 (0.70)                    & 27.20    \\
                                    & GHS                  & 2.45 (0.12)                     & 57.72 (5.35)                     & 0.20 (0.08)                     & 20.35    \\
                                    & quasiGHS-diag         & 2.13 (0.08)                     & 56.60 (4.99)                     & 0.26 (0.10)                     & 7.77     \\
                                    & quasiGHS              & 2.69 (0.14)                     & 56.74 (5.61)                     & 0.31 (0.12)                     & 7.91     \\ \hline
\multirow{6}{*}{Hubs   + Cliques}   & GL                   & 2.67 (0.27)                     & 83.83 (15.24)                    & 6.51 (4.53)                     & 0.06     \\
                                    & GSCAD                & 2.97 (0.04)                     & 50.43 (7.91)                     & 0.25 (0.16)                     & 0.09     \\
                                    & BGL                  & 3.83 (0.11)                     & 96.37 (2.15)                     & 19.81 (0.63)                    & 26.89    \\
                                    & GHS                  & 2.10 (0.10)                     & 54.30 (3.81)                     & 0.13 (0.06)                     & 19.87    \\
                                    & quasiGHS-diag         & 2.53 (0.07)                     & 61.13 (6.14)                     & 0.38 (0.12)                     & 7.38     \\
                                    & quasiGHS              & 2.28 (0.14)                     & 59.10 (5.46)                     & 0.24 (0.10)                     & 7.97     \\
\hline\hline
\end{tabular}
\end{table}

\begin{figure}[ht]
    \centering
    \includegraphics[width=5in, height=3.5in]{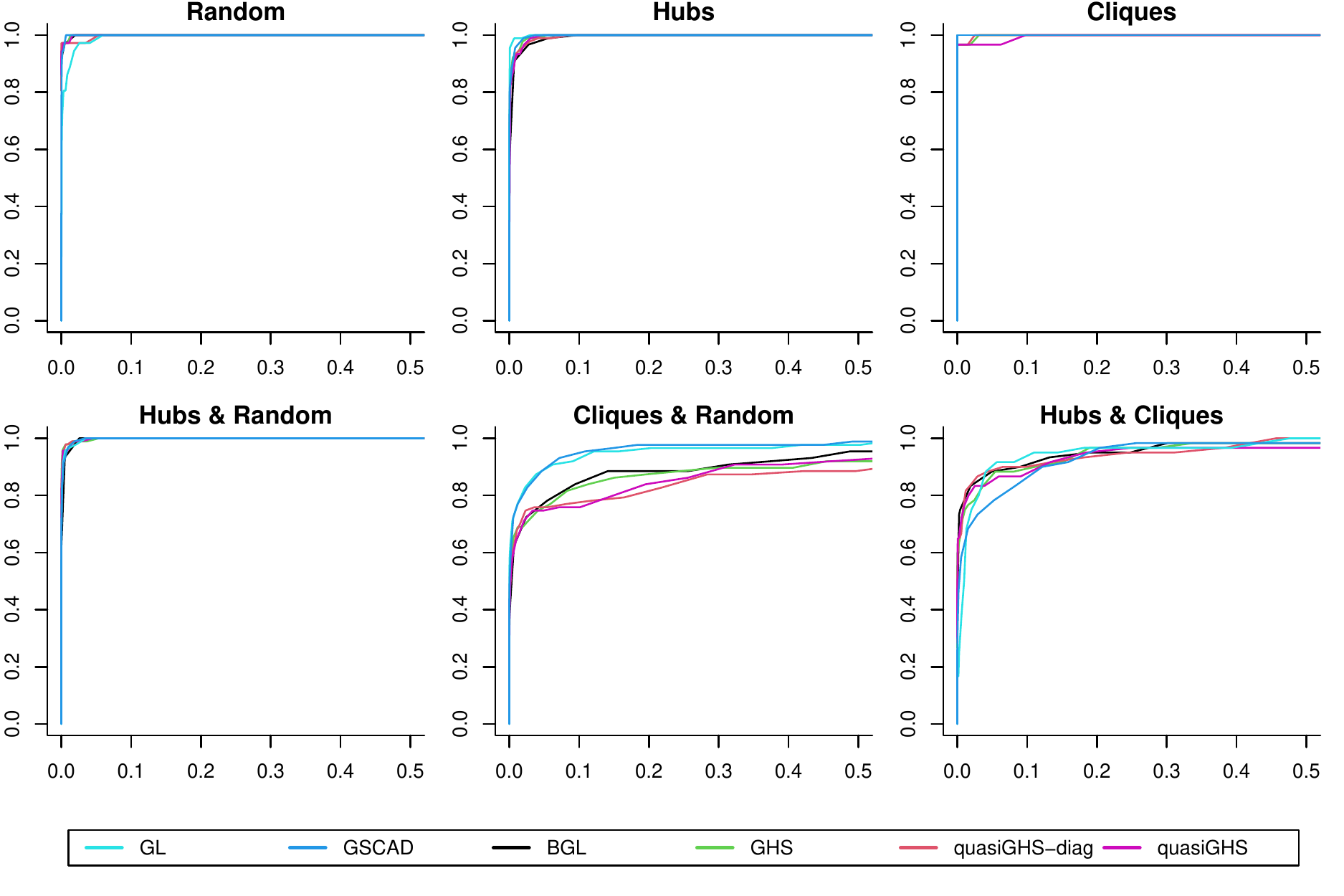}
    \caption{ROC Curve when p=100, n=150. x-axis is FPR and y-axis is TPR. \label{fig:roc}}
\end{figure}

From Table \ref{tbl:p100n150}, we can see that quasiGHS-diag has the smallest error rate with respect to Frobenius norm in the first five patterns, which is not surprising because the true diagonal values all known in quasiGHS-diag. Yet the performance of quasiGHS is comparable to that of quasiGHS-diag. Compared with GHS which uses the full likelihood function, quasiGHS has only slightly higher error rate. Moreover, judging from the FPRs and TPRs, quasiGHS has similar variable selection performance as GHS and GSCAD do.
Compared with the frequentist methods (GL, GSCAD), quasiGHS has the advantage of quantifying the uncertainty by providing credible regions. In addition, the computation complexity of quasiGHS is much lower than BGL and GHS. This advantage will be more significant as $p$ increases. 

Figure \ref{fig:roc} shows the ROC curves of GL, GSCAD, BGL, GHS, quasiGHS-diag, and quasiGHS, respectively on one randomly chosen data set out of $50$. The ROC curves for GL and GSCAD are generated by varying tuning parameters $\lambda$. The ROC curves for BGL, GHS, quasiGHS-diag and quasiGHS are obtained by increasing the length of the posterior credible intervals from $1\%$ to $99.99\%$. Except for the Cliques+Random pattern, the six methods exhibit similar ROC curves. In the Clique+Random pattern, frequentist methods GL and GSCAD outperform all Bayesian methods, and quasiGHS has similar performance as GHS does.

\section{An Illustration with Real Data}
We now apply the quaisGHS method to analyze a real world data set, CEU. The data set CEU records the expression levels of 100 human genes in 60 unrelated individuals of Northern and Western European ancestry from Utah. The detailed description for this data set can be found in \cite{bhadra2013joint}. It has been studied by \cite{bhadra2013joint} in a regression model where the SNPs are used as regressors and the gene expressions as response variables. A network was plotted based on the estimated residual precision matrix. Later \cite{li2019graphical} analyzed the data set under a mean zero Gaussian graphical model. 

Here, we use the same setting as in \cite{li2019graphical} and compared the performance of quasiGHS with those of GL, GSCAD, BGL, and GHS. For GL and GSCAD, each nonzero off-diagonal element $\hat\omega_{ij}$ in the estimated precision matrix indicates the existence of association between the expressions of a gene pair $(i,j)$. For BGL, GHS and quasiGHS, if the $50\%$ posterior credible interval of $\omega_{ij}$ does not include zero, we would draw an edge between the gene $i$ and gene $j$.  

\begin{table}[!ht]
\centering
\caption{Number of vertices and edges estimated \label{tbl:CEU_est}}
\begin{tabular}{lccccc}
\hline\hline
            & GL   & GSCAD & BGL & GHS & quasiGHS \\ \hline
\# vertices & 100  & 94    & 100 & 82  & 88      \\ \hline
\# edges    & 1087 & 361   & 980 & 105 & 129    \\ \hline\hline 
\end{tabular}
\end{table}

\begin{figure}[ht]
\centering
\begin{subfigure}{.32\textwidth}
  \centering
  \includegraphics[width=1\linewidth]{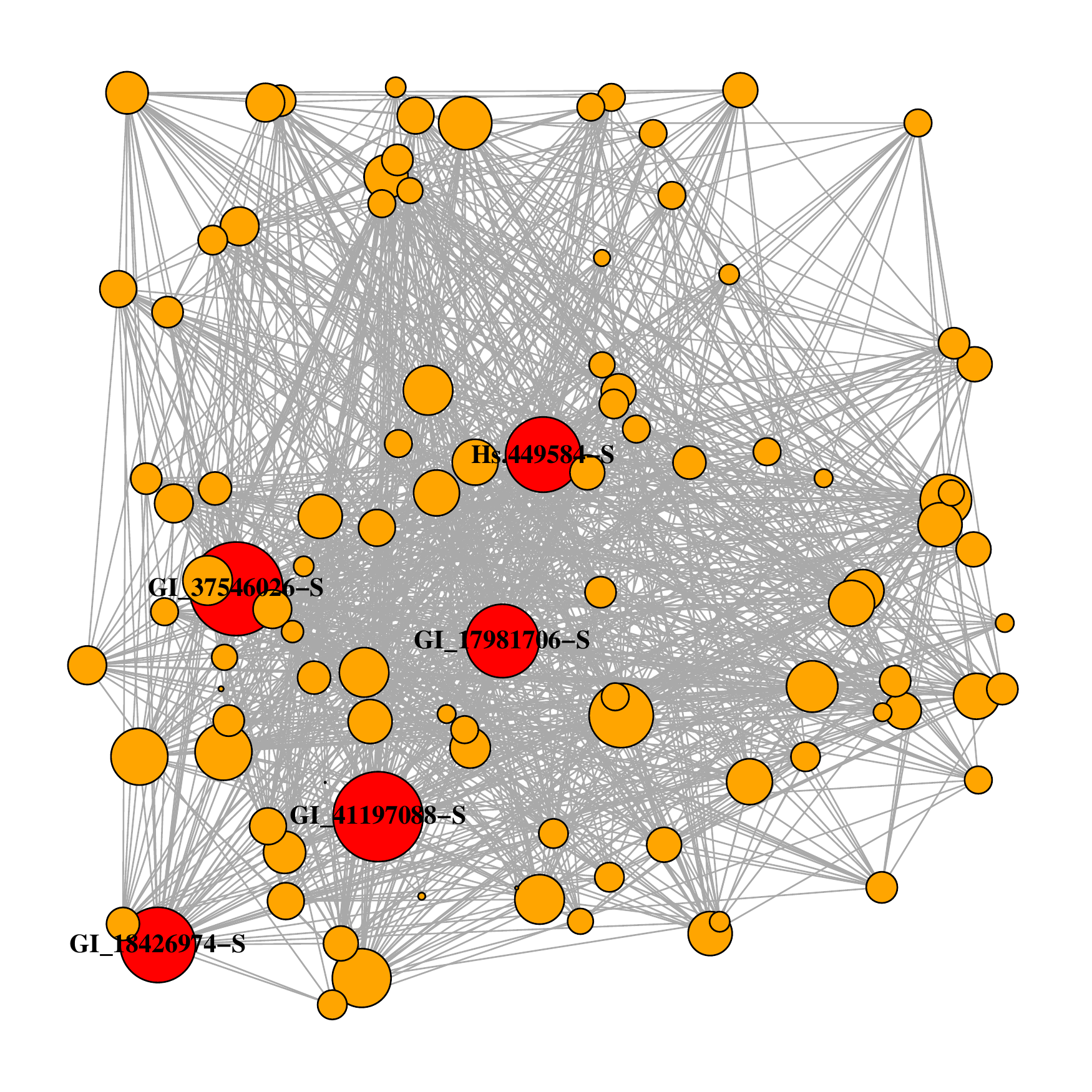}  
  \caption{graphical Lasso}
  \label{fig:sub-gLasso}
\end{subfigure}
\hfill
\begin{subfigure}{.32\textwidth}
  \centering
  \includegraphics[width=1\linewidth]{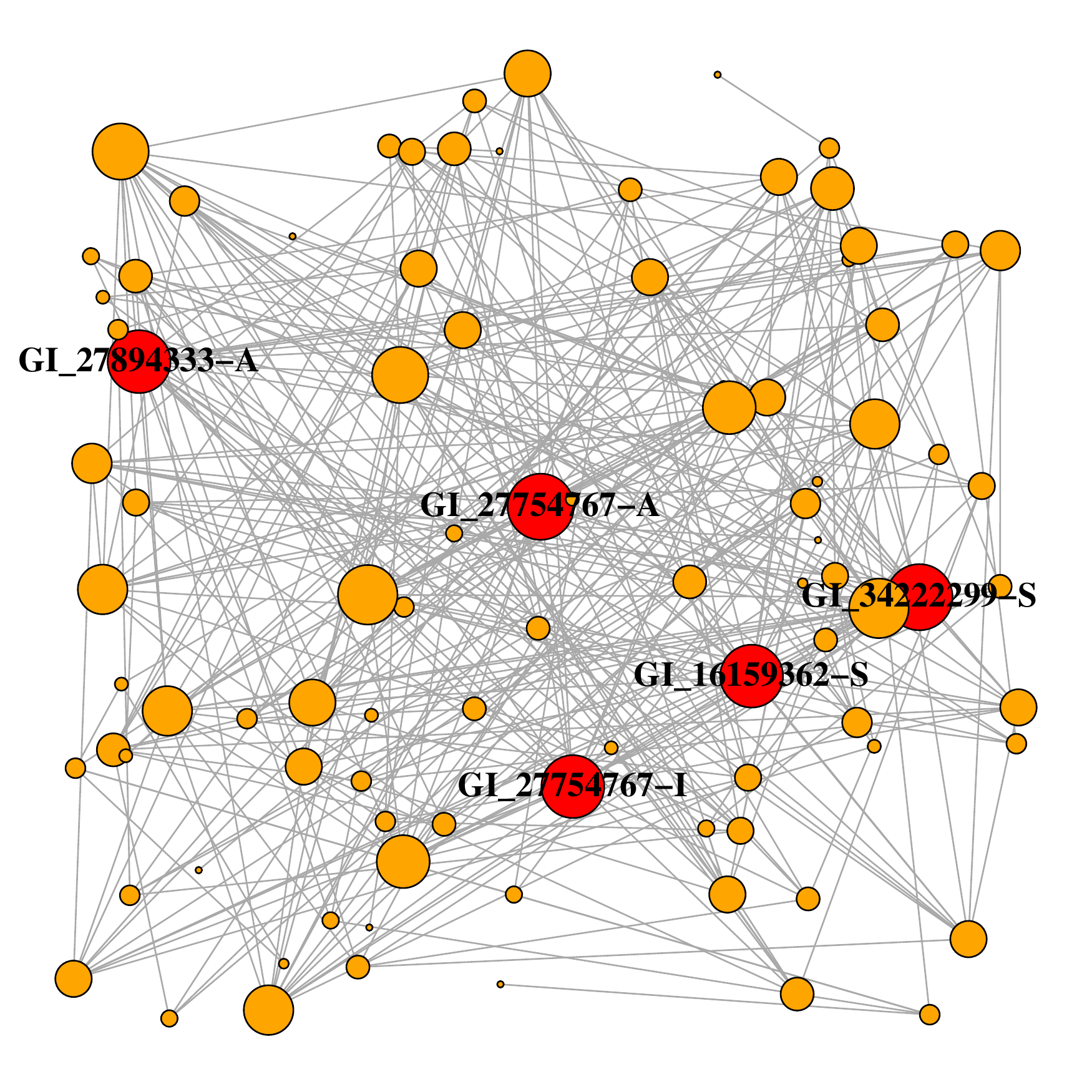}  
  \caption{graphical SCAD}
  \label{fig:sub-gSCAD}
\end{subfigure}

\begin{subfigure}{.32\textwidth}
  \centering
  \includegraphics[width=1\linewidth]{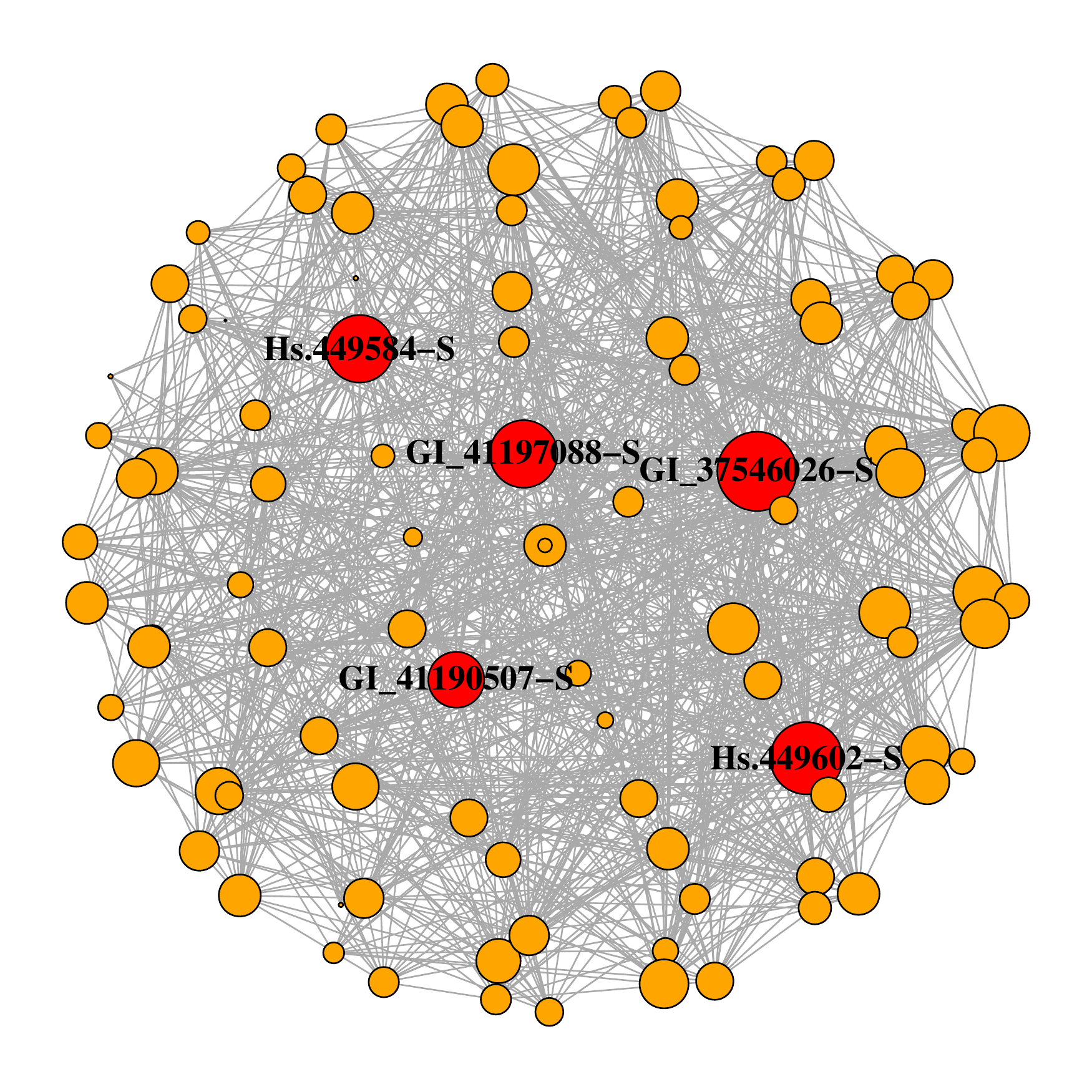}  
  \caption{BGL}
  \label{fig:sub-BGL}
\end{subfigure}
\hfill
\begin{subfigure}{.32\textwidth}
  \centering
  \includegraphics[width=1\linewidth]{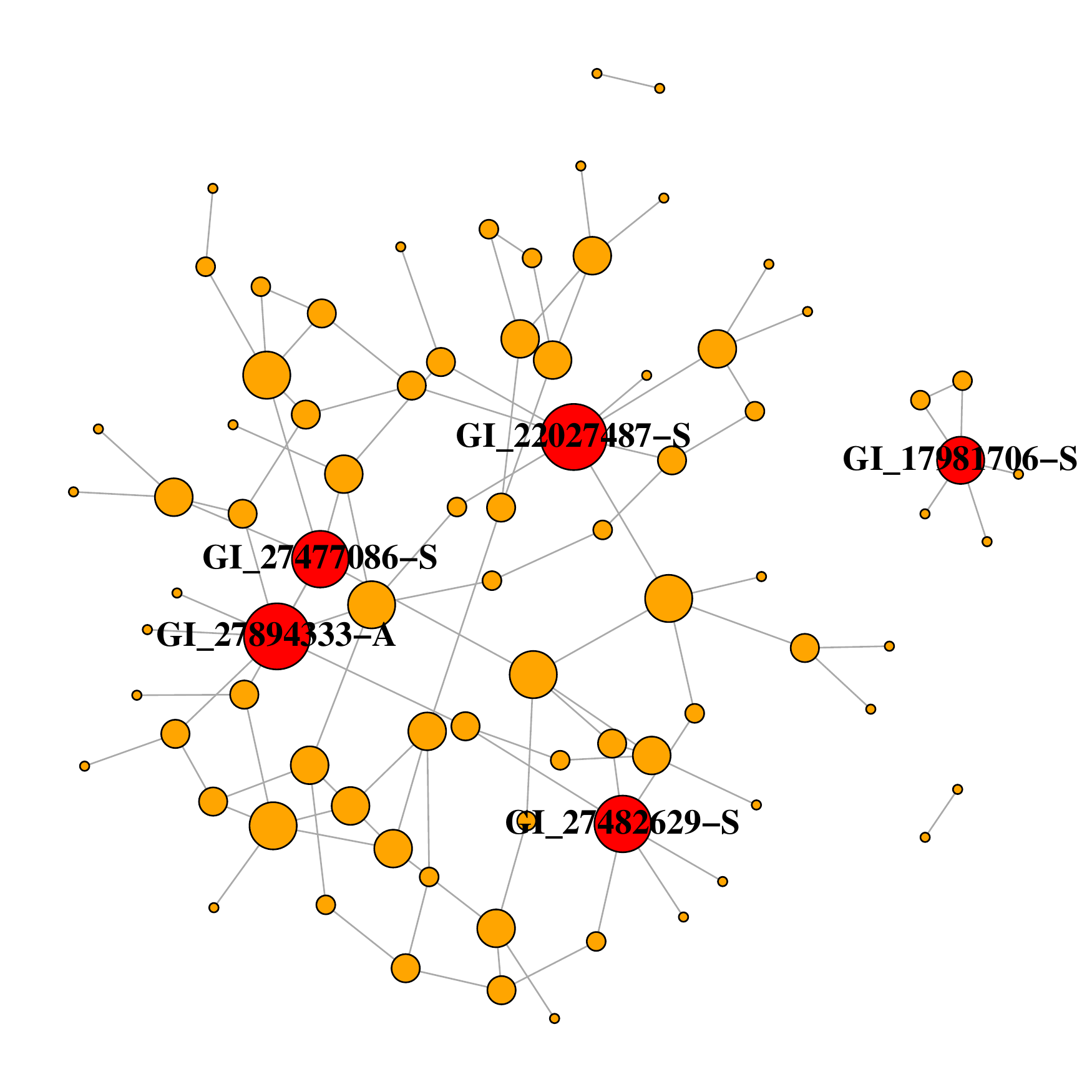}  
  \caption{GHS}
  \label{fig:sub-GHS}
\end{subfigure}
\hfill
\begin{subfigure}{.32\textwidth}
  \centering
  \includegraphics[width=1\linewidth]{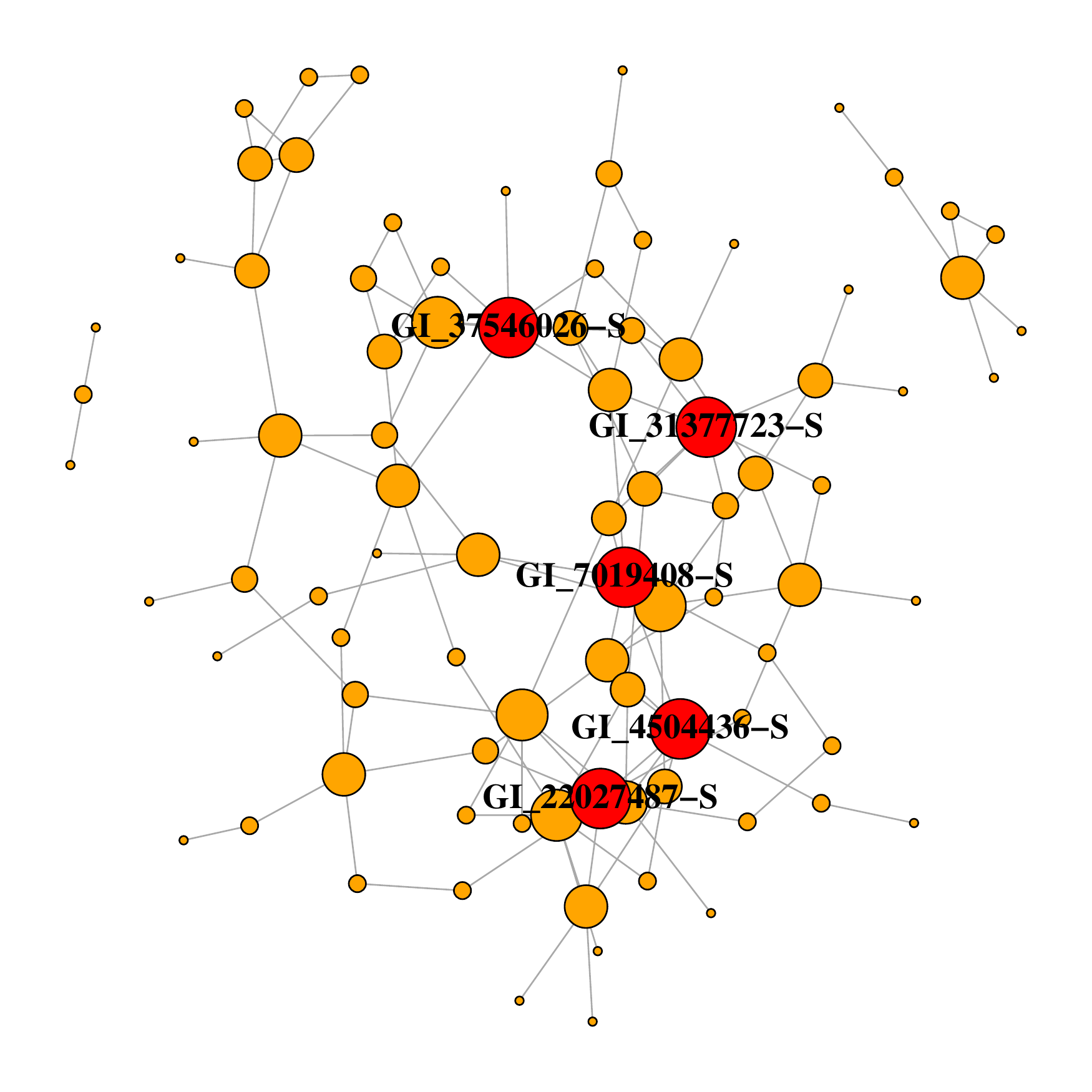}  
  \caption{quasiGHS}
  \label{fig:sub-quasiGHS}
\end{subfigure}
\hfill
\caption{Inferred graphs for the CEU data by GL, GSCAD, BGL GHS, and quasiGHS. Size of node is proportional to degree within each graph.}
\label{fig:CEU_6fig}
\end{figure}

Table \ref{tbl:CEU_est} listed the number of edges and the number of connected vertices estimated by the five methods. A connected vertex means a node with at least one edge. The corresponding inferred graphs are displayed in Figure \ref{fig:CEU_6fig}. We can see that GHS provides the sparsest graph, and the second sparsest graph is given by quasiGHS. The other methods give much denser graphs. The five genes with largest degrees (the number of edges connected) are marked out in red. Nodes that are not connected with any other genes are not shown in plot.

\section{Conclusion and Future Work}
In this work, we proposed a new method, the quasi-Bayesian framework with a class of general shrinkage priors, for sparse precision matrix estimation and variable selection.
The continuous shrinkage priors need to have both high concentration around zero and heavy tail. The method proposed can obtain optimal posterior contraction rate comparable with that of the spike-and-slab prior with multivariate Gaussian likelihood, in terms of the Frobenius norm and spectral norm of estimation error. Simulations confirm the good performance of quasiGHS on estimation and variable selection in various settings. And quasiGHS also shows computational efficiency compared with methods using the full likelihood function, like GHS and BGL. 

There remains several open problems following this work.
In the derivation of the theoretical results, we assume the diagonal elements to be known. 
While the simulations show similar performances in the both cases of known and unknown diagonal elements, it is worth further efforts to establish the posterior consistency when the diagonal elements are unknown. Another aspect to work on is to accelerate the computation by parallel computing. 
\cite{terenin2019gpu} demonstrated with a Horseshoe Probit regression model that Gibbs sampling in a fully data-parallel manner on a graphics processing unit (GPU) could reduce the computation cost by a large amount. Developing a fast computing tool with parallel mechanisms would help advance the method proposed. 
Finally, there are several types of pseudo-likelihood functions other than \eqref{eq:pseudo-llh} that could be employed, like the pseudo-likelihoods used by CONCORD \citep{khare2015convex} and SPLICE \citep{rocha2008path}. Whether we could establish similar theoretical properties for other pseudo-likelihood based schemes requires further exploration.

\bibliographystyle{apalike}
\bibliography{Pseudo_GGM.bib}

\section*{Appendix}
\begin{lemma} \label{lemma:song}
(Lemma A.3 in \cite{song2020bayesian}) Let $\fstar$ be the true probability density of data generation, $\htheta$ be a likelihood function with parameter $\theta \in \Theta$, where $\Theta$ is the parameter space. $\Estar$ and $\Etheta$ denote the expectation w.r.t. $\fstar$ and $\htheta$ respectively. Let $\setBn$ and $\setCn$ be two subsets of $\Theta$, and $\phi$ be some testing function satisfying $\phi(\bZ) \in [0, 1]$ for data $\bZ$.\\
If $\pi(\setBn)\leq b_n$, $\E^*[\phi(\bZ)]\leq b_n'$, $\sup_{\theta\in \setCn} \Etheta[1-\phi(\bZ)]\leq c_n$, and 
\[
\Pstar\left(\dfrac{m(\bZ)}{\fstar(\bZ)} \geq a_n \right) \geq 1-a_n',
\]
where $m(\bZ)=\int_\Theta\pi(\theta)\htheta(\bZ)d\theta$, and $\Pstar$ denotes the probability under $\fstar$, then,
\[
\Estar[\pi_h(\setCn \cup B_n)|\bZ] \leq \dfrac{b_n+c_n}{a_n} + a_n' + b_n',
\]
where $\pi_h(\setA)=\int_\setA \pi(\theta)\htheta(\bZ)d\theta / m(\bZ)$ for $\setA \subset \Theta$.
\end{lemma}

\begin{lemma} \label{lemma:inequalities}
For any Gaussian random design $X \in \Rnp$ with i.i.d. $N(0, \Sigma)$ rows. Define $\bar{\sigma}=\sqrt{\max_{j=1,...,p}\Sigma_{jj}}$. Then we have the following three properties for $X$:

(1) There are universal positive constants $c$, $c'$ such that 
\begin{equation}\label{raskutti}
    \dfrac{\normtwo{Xv}}{\sqrt{n}} \geq \dfrac{1}{4}\normtwo{\Sigma^{1/2}v} - 9\bar{\sigma}\sqrt{\dfrac{\log p}{n}}\normone{v} \quad \text{for all } v\in\Rp,
\end{equation}
with probability at least $1-c'\exp(-cn)$.\\

(2) There are universal positive constants $\tilde{c}$ and $\tilde{c}'$ such that
\begin{equation}\label{vershynin}
    \lambda_{\max}(X^TX) \leq (2\sqrt{n} + \tilde{c}\sqrt{p})^2
\end{equation}
with probability at least $1-2\exp(-\tilde{c}'n)$.\\

(3) Let $X_{.i}$ denote the $i$-th column of matrix $X$, then with probability at least $1 - p\exp(-n/8)$
\begin{equation} \label{maxXi2}
    \max_i \normtwo{X_{.i}}^2 \leq 2\bar{\sigma}^2n.
\end{equation}
\end{lemma}

\begin{proof}
Part (1) is Theorem 1 in \cite{raskutti2010restricted}. Part (2) can be obtained by setting $t=\sqrt{n}$ in Theorem 5.39 of \cite{vershynin2010introduction}.

For Part (3), since $\normtwo{X_{.i}}^2/\Sigma_{ii} \sim \chi^2(n)$, by concentration inequalities of chi-square distribution, we have 
\[
\mathbb{P}(\normtwo{X_{.i}}^2/\Sigma_{ii}>2n) \leq \exp(-\frac{n}{2}(2-1-\log 2)) < \exp(-n/8).
\]
And $\mathbb{P}(\max_i \normtwo{X_{.i}}^2 > 2\bar{\sigma}^2n) \leq p\cdot\mathbb{P}(\normtwo{X_{.i}}^2/\Sigma_{ii}>2n) \leq p\exp(-n/8).$
\end{proof}

Note that by Borel-Cantelli Lemma, all the three inequalities \eqref{raskutti}, \eqref{vershynin} and \eqref{maxXi2} hold almost surely since $\log p \simeq \log n$. We will refer to the three inequalities later without repeatedly stating the almost sure convergence results. 

\begin{lemma}[Prior Concentration] \label{lemma:prior_con} Consider the prior $\pialpha(\bm{\theta})=\prod_{i=1}^p\pialpha(\theta_i)$, where $\pialpha(\theta_i)$ satisfies (\ref{cond:prior}).
Let 
$ \setBn:=\{\theta\in\Rp: \sum_{j=1}^p I(|\theta_j|>a_n) > L\sstar\}$, then $\pi_\alpha(\setBn) \leq \exp(-uLn\epsilon_n^2)$, where $L$ is a positive constant free of $n$ and $p$. 

\end{lemma}

\begin{proof}
  Let $v:=\int_{|\theta_j| \geq a_n} \pi(\theta)d\theta \leq p^{-(1+u)}$, by Lemma A.3 in \cite{song2017nearly}, for $1<L\sstar<p-1$
 \begin{equation*}
     \pi(\setBn) \leq 1-\Phi(\sqrt{2p \cdot H(v, (L\sstar)/p}),
 \end{equation*}
 where $\Phi$ is the CDF of standard Gaussian distribution and 
 $H(v,k/p)=(k/p)\log(k/pv)+(1-k/p)\log[(1-k/p)/(1-v)]$. 
 
Note that $
     p \cdot H(v,\frac{L\sstar}{p}) 
     = (L\sstar)\log(\frac{L\sstar}{pv}) +
     (p-L\sstar)\log(\frac{p-L\sstar}{p-pv})
     \geq uL\sstar\log p
     \geq uL n\epsilon_n^2$.

Therefore, for large enough $p$, 
 \begin{align} \label{eq:prior_concentration}
      \pi(\setBn) 
      \leq \dfrac{\exp\{-p \cdot H(v, L\sstar/p\}}{\sqrt{2\pi}\sqrt{2p \cdot H(v, L\sstar/p)}}
      \leq \dfrac{\exp\{-uL n\epsilon_n^2\}}{2\sqrt{\pi uL n\epsilon_n^2}}
      \leq \exp\{-uL n\epsilon_n^2\}.
 \end{align}
\end{proof}

\begin{lemma}[Existence of test]\label{lemma:existtest}
Consider the linear regression model \eqref{eq:linearreg}. Suppose $n$, $p$ and the true $\thetastar$ satisfy conditions $C_1$. Then for design matrix $X$ satisfies inequalities \eqref{raskutti} and \eqref{maxXi2} in Lemma \ref{lemma:inequalities}, there exists a measurable function $\phi:\mathcal{Z}^{(n)} \to [0,1]$, such that 
\[\Estar[\phi(\bZ)] \leq 4/p^2,\]
\[\sup_{\theta\in \setCn} \Etheta[1-\phi(\bZ)] \leq \exp(-c_2n\epsilon_n^2),\]
where $c_2=\lambda_{\min}(\Sigma)M^2/256$ for some constant $M$.
\end{lemma}

\begin{proof} The proof shares similar ideas as \cite{atchade2017contraction, wei2020contraction}, but we did not work on an uniformly bounded design matrix. \\

Let $\nabla \log \fstar(\bZ):=\dfrac{d\log f_\theta(\bZ)}{d\theta}|_{\theta=\thetastar}$. Define $\sbar=(L+1)\sstar$. For a non-empty set $\setTilTheta:=\{\theta\in\Rp: \sum_{j=1}^p I(|\theta_j|>a_n) \leq \sbar\}$, we define a subset $\setTilEps$ of the sample space $\splspcZn$ by
\[
\setTilEps:=\{ \bZ \in \splspcZn: \text{for all } \theta \in \thetastar + \setTilTheta, \;
  \norminf{\nabla \log \fstar(\bZ)} \leq \lambda/2 \}.
\]
The value of $\lambda$ will be determined later.

Define $\setBallp(\setTilTheta, M\epsilon_n) := \{\theta \in \thetastar + \setTilTheta: \normtwo{\theta - \thetastar} \leq M\epsilon_n \} = \left( \bigcup_{l\geq 1} B(l) \right)^c$, where $B(l):=\{\theta \in \thetastar + \setTilTheta: l M\epsilon_n \leq \normtwo{\theta - \thetastar} \leq (l+1)M\epsilon_n \}$. \\

Let $\setSl$ be the set of maximal points separated by the distance $lM\epsilon_n/2$ in $B(l)$. Define a function $q_\theta(\bZ):=I(\bZ\in\setTilEps)\cdot f_\theta(\bZ)$. For each $\thetakl \in \setSl$, let $\setQthetak:=\{q_u: u\in \thetastar + \setTilTheta \text{ and } \normtwo{u-\thetakl} \leq lM\epsilon_n/2\}$. Denote the convex hull of the set $\setQthetak$ by $\setConvQ$. \\

By Lemma 12 in \cite{atchade2017contraction}, with $p=\fstar$ and $q \in \setConvQ$, there exists a test function $\phi_{\thetakl}: \splspcZn \to [0,1]$ such that
\[
\Estar[\phi_{\thetakl}(\bZ)] \leq \sup_{q \in \setConvQ}\Hhalf(\fstar,q)
\]
\[
\sup_{q \in \setConvQ} \int_{\splspcZn} (1-\phi_{\thetakl}(z))q(z)dz \leq \sup_{q\in \setConvQ}\Hhalf(\fstar,q),
\]
where $\Hhalf(p_1, p_2):=\int_{\splspcZn}\sqrt{p_1(z)p_2(z)}dz$.\\

Since $q \in \setConvQ$, it can be written as a finite convex combination $q=\sum_i \alpha_i q_{u_i}$, where $\sum_i \alpha_i=1$, $u_i \in \thetastar + \setTilTheta$ and $\normtwo{u_i - \thetakl} \leq lM\epsilon_n/2$. And note that $\normtwo{\thetakl-\thetastar}\geq lM\epsilon_n$ because $\thetakl \in B(l)$. So we have $\normtwo{u_i-\thetastar}\geq lM\epsilon_n/2 > \epsilon_n$ for $M>2$.\\

Note that 
\[
\Hhalf(\fstar, q) = \int_{\splspcZn} \fstar(z) \left( \sum_i \alpha_i\dfrac{  q_{u_i}(z)}{\fstar(z)} \right)^{1/2} dz,
\]
we first establish the upper bound for $q_{u_i}(z)/\fstar(z)$.

Note that
\begin{align*}
    \dfrac{q_{u_i}(z)}{\fstar(z)} 
    &= I(z \in\setTilEps) \dfrac{f_{u_i}(z)}{\fstar(z)} \\
    &= I(z \in\setTilEps) \exp\{\nabla\log\fstar(z)(u_i-\thetastar) + L_{u_i}(z)\}
\end{align*}
where $L_{u_i}(z) = -(u_i-\thetastar)^TX^TX(u_i-\thetastar)/2$.

Since $u_i-\thetastar \in \setTilTheta$ and $\normtwo{u_i-\thetastar} > \epsilon_n$, $
\normone{u_i-\thetastar} \leq 2\sqrt{\sbar}\normtwo{u_i-\thetastar}
$. By inequality \eqref{raskutti}, for  $n\geq 144\bar{\sigma}^2\sbar\log p/\lambda_{\min}(\Sigma)$, 
\begin{align*}
    \dfrac{\normtwo{X(u_i-\thetastar)}}{\sqrt{n}} 
    &\geq \dfrac{1}{4}\normtwo{\Sigma^{1/2}(u_i-\thetastar)} - 18\bar{\sigma}\sqrt{\dfrac{\sbar\log p}{n}}\normtwo{u_i-\thetastar}\\
    &\geq \dfrac{1}{8}\lambda_{\min}^{1/2}(\Sigma)\normtwo{u_i-\thetastar}
\end{align*}
where $\lambda_{\min}(\Sigma)$ is the minimum eigenvalue of $\Sigma$. Hence,
$L_{u_i}(z) \leq -\dfrac{1}{16}\lambda_{\min}(\Sigma) n\normtwo{u_i-\thetastar}^2$.

For $z\in \setTilEps$, $\norminf{\nabla\log\fstar(z)}\leq \lambda/2$ and
\[
\nabla\log\fstar(z)(u_i-\thetastar)  \leq \norminf{\nabla\log\fstar(z)}\normone{u_i-\thetastar} \leq \lambda\sqrt{\sbar}\normtwo{u_i-\thetastar},
\]
so we have
\begin{align*}
\dfrac{q_{u_i}(z)}{\fstar(z)} &\leq \exp\{\lambda\sqrt{\sbar}\normtwo{u_i-\thetastar}  -\dfrac{1}{16}\lambda_{\min}(\Sigma) n\normtwo{u_i-\thetastar}^2\}\\
& \leq \exp\{\left(\lambda\sqrt{\sbar}/\normtwo{u_i-\thetastar} -\dfrac{n}{16}\lambda_{\min}(\Sigma)\right)\normtwo{u_i-\thetastar}^2 \} \\
& \leq \exp\{-\dfrac{\lambda_{\min}(\Sigma)M^2}{128}l^2n\epsilon_n^2 \}
\end{align*}
if
$ \dfrac{\lambda\sqrt{\sbar}}{\normtwo{u_i-\thetastar}} \leq 
\dfrac{2\lambda\sqrt{\sbar}}{M\epsilon_n} \leq \dfrac{n}{32}\lambda_{\min}(\Sigma) $, 
\textit{i.e.}, 
$ \lambda \leq  \dfrac{\lambda_{\min}(\Sigma)}{64\sqrt{L+1}}M\sqrt{n\log p}$. Hence, 
\[ \sup_{q \in \setConvQ}\Hhalf(\fstar, q) \leq \exp\{-c_1M^2ln\epsilon_n^2\}, \quad c_1 = \lambda_{\min}(\Sigma)/256. \]

Uptil now, we have shown that 
\begin{align}
\Estar[\phi_{\thetakl}(\bZ)] &\leq \exp\{-c_1M^2ln\epsilon_n^2\} \label{existtest_prep:1} \\ 
\sup_{q \in \setConvQ} \int_{\splspcZn} (1-\phi_{\thetakl}(z))q(z)dz & \leq \exp\{-c_1M^2ln\epsilon_n^2\} \label{existtest_prep:2}
\end{align}

Let $ \phi(z)=I(z\in\setTilEps)\cdot\sup_{l\geq 1}\max_{\thetakl\in\setSl} \phi_{\thetakl} + I(z\in\setTilEps^c)$.

Consider $\setCn= \{\theta\in\thetastar+\setTilTheta: \normtwo{\theta-\thetastar}>M\epsilon_n\}$. Then for $\theta\in \setCn$, since $\theta\in B(l)$ for some $l\geq 1$, it is within the distance of $M\epsilon_n/2$ of a point $\thetakl\in\setSl$, $l\geq 1$, which implies $q_\theta \in \setQthetak$.

By (\ref{existtest_prep:2}), 
\begin{equation} \label{existtest:2}
    \sup_{\theta\in \setCn} \Etheta[1-\phi(Z)] \leq \exp\{-c_1M^2n\epsilon_n^2\}.
\end{equation}

Let $D_l:=D(lM\epsilon_n, \setBallp(\setTilTheta, (l+1)M\epsilon_n))$, where $D(\epsilon, \mathcal{B})$ denotes the $\epsilon$-packing number of the set $\mathcal{B}$. By \cite{wei2020contraction}
\begin{equation} \label{existtest:1_1}
     \Estar[\phi(Z)] \leq \Pstar(Z\in \setTilEps^c) + \sum_{l\geq 1}D_l\exp\{-c_1M^2ln\epsilon_n^2\},
\end{equation}
and $D_l \leq \binom{p}{\sbar} (24)^{\sbar} \leq 24^{\sbar} e^{\sbar\log(p e)}$ \citep{ghosal2000convergence}. Hence, 
\begin{align} \label{existtest:1_2}
    \begin{split}
        \sum_{l\geq 1}D_l\exp\{-c_1M^2ln\epsilon_n^2\}
        & \leq 2\exp\{\sbar\log(24e)+\sbar\log p - c_1M^2n\epsilon_n^2\}\\
        & \leq 2\exp\{-2\sbar\log p\} \text{ if }  M^2>8L/c_1 \\
        & \leq \dfrac{2}{p^2}.
    \end{split}
\end{align}

Next we will show that $\Pstar(Z\in \setTilEps^c)\leq 2/p^2$ for properly chosen $\lambda$. 
\begin{align} \label{existtest:1_3}
    \begin{split}
        \Pstar(Z\in \setTilEps^c) 
        & = \Pstar( \norminf{\nabla \log \fstar(\bZ)} > \lambda/2) \\
        & = \Pstar( \norminf{X^T(Z-X\thetastar)} > \lambda/2) \\
        & = \P( \norminf{Y} > \lambda/2) \text{ where } Y \sim N(0, X^TX) \\
        & \leq \sum_{i=1}^p \P(|Y_i|>\lambda/2) \text{ where } Y_i \sim N(0, \normtwo{X_{.i}^2}) \\
        & \leq \sum_{i=1}^p 2\exp\left(-\dfrac{\lambda^2}{8\normtwo{X_{.i}}^2} \right) \\
        & \leq 2\exp\left(-\dfrac{\lambda^2}{8\max_i\normtwo{X_{.i}}^2} + \log p\right) \\
        & \leq \dfrac{2}{p^2}  \text{ if }  \lambda\geq \sqrt{24\max_i\normtwo{X_{.i}}^2 \cdot \log p}.
    \end{split}
\end{align}

By the steps above, we see that 
\[ \sqrt{24\max_i\normtwo{X_{.i}}^2 \cdot \log p} \leq \lambda \leq  \dfrac{\lambda_{\min}(\Sigma)}{64\sqrt{L+1}}M\sqrt{n\log p}. \]
Note that $\max_i \normtwo{X_{.i}}^2 \leq 2\bar{\sigma}^2n$ because $\bX$ satisfies inequality \eqref{maxXi2}. So such $\lambda$ will exist with sufficiently large constant $M$. 

Combining \eqref{existtest:1_1}, \eqref{existtest:1_2} and \eqref{existtest:1_3}, we have
\begin{equation} \label{existtest:1}
    \Estar[\phi(\bZ)] \leq 4/p_n^2.
\end{equation}
\eqref{existtest:1} and \eqref{existtest:2} complete the proof of Lemma \ref{lemma:existtest}.

\end{proof}

\begin{lemma}\label{lemma:marginal} Consider the linear regression model \eqref{eq:linearreg}. Suppose $n$, $p$ and the true $\thetastar$ satisfy conditions $C_1$, and $\bX$ satisfies inequality \eqref{vershynin} in Lemma 2. The prior is given by $\pialpha(\theta)=\prod_{i=1}^p\pialpha(\theta_i)$, where $\pialpha(\theta_i)$ satisfies (\ref{cond:prior}). Then the following inequality holds with some positive constant $c_4$:
\[
\Pstar\left(\int_{\Rp} \dfrac{f_{\bm\theta}(\bZ)}{f_{\thetastar}(\bZ)}\pialpha(\bm{\theta})d\bm{\theta} \geq \exp(-c_4 n\epsilon_n^2) \right) > 1-\dfrac{2}{p^2}.
\]
\end{lemma}
\begin{proof}
Note that
\begin{equation*}
    \int_{\Rp} \dfrac{f_{\bm\theta}(\bZ)}{f_{\thetastar}(\bZ)}\pialpha(\bm{\theta})d\bm{\theta}
    = \int_{\Rp} \exp\{\nabla \log f_{\thetastar}(\bZ)^T(\bm\theta-\thetastar) -\dfrac{1}{2}\normtwo{\bX(\bm\theta-\thetastar)}^2 \} \pi(\bm\theta) d\bm\theta.
\end{equation*}

Since $\bX$ satisfies \eqref{vershynin} and $p<k_1^2 n$, for some constant $k_1 > 0$,
\[
\normtwo{\bX(\bm\theta-\thetastar)}^2 \leq (\ck+2)^2n\normone{\bm\theta-\thetastar}^2.
\]

We fist consider $z \in \setEps:=\{\bZ\in\splspcZn:  \norminf{\nabla \log f_{\thetastar}(\bZ)} \leq \lambda/2 \}$, where $\lambda$ satisfies
\[ \sqrt{24\max_i\normtwo{X_{.i}}^2 \cdot \log p} \leq \lambda \leq  \dfrac{\lambda_{\min}(\Sigma)}{64\sqrt{L+1}}M\sqrt{n\log p} \]
as in Lemma \ref{lemma:existtest}.

Note that $
\nabla \log f_{\thetastar}(z)^T(\bm\theta-\thetastar) \leq 
\norminf{\nabla\log f_{\thetastar}(z)}\normone{\bm\theta-\thetastar}
\leq \dfrac{\lambda}{2}\normone{\bm\theta-\thetastar}$, Hence, for some constant $\eta>0$, we have
\begin{align*}
    \int_{\Rp}\dfrac{f_{\bm\theta}(z)}{f_{\thetastar}(z)}\pi(\bm\theta)d\bm\theta 
    &\geq \int_{\Rp} \exp\{-\dfrac{\lambda}{2}\normone{\bm\theta-\thetastar}- 
    \dfrac{(\ck+2)^2}{2}n\normone{\bm\theta-\thetastar}^2 \} \pi(\bm\theta)d\bm\theta \\
    &\geq \exp\{-\lambda\eta\epsilon_n - 2(\ck+2)^2\eta^2n\epsilon_n^2 \}
    \cdot \pi(\normone{\bm\theta-\thetastar}<2\eta\epsilon_n).
\end{align*}

Since $\lambda\leq 4c_1M\sqrt{n\log p/(L+1)}=4c_1Mn\epsilon_n/\sqrt{\sbar}$, we have
\[ 
\exp\{-\lambda\eta\epsilon_n - 2(\ck+2)^2\eta^2n\epsilon_n^2\} \geq
\exp\{-c_3 n\epsilon_n^2\}
\]
where $c_3 = \dfrac{4c_1M\eta}{\sqrt{\sbar}} +  2(\ck+2)^2\eta^2$.

Now we will look at $\pi(\normone{\bm\theta-\thetastar}<2\eta\epsilon_n)$.
Note that
\begin{align*}
    \pi(\normone{\bm\theta-\thetastar}<2\eta\epsilon_n) \geq 
     \prod_{j\in\xistar} \pi \left(\theta_j\in [\thetastar_j-\frac{\eta\epsilon_n}{\sstar}, \thetastar_j+\frac{\eta\epsilon_n}{\sstar}] \right)
    \times   \prod_{j\notin\xistar} \pi \left( |\theta_j| \leq a_n \right)
\end{align*}
and
\[
\pi \left(\theta_j\in [\thetastar_j-\frac{\eta\epsilon_n}{\sstar}, \thetastar_j+\frac{\eta\epsilon_n}{\sstar}] \right)
\geq \dfrac{2\eta\epsilon_n}{\sstar}\cdot \inf_{\theta_j \in [-E_n, E_n]} \pi(\theta_j).
\]

By \eqref{cond:fattail}, $\inf_{\theta_j \in [-E_n, E_n]} \pi(\theta_j) \geq p^{-c}$. And $n\leq p^k$, so we have
\[ \epsilon_n/\sstar = (\log p/(n\sstar))^{1/2}\geq p^{-(k+1)/2}.\] 

Combining $\prod_{j\notin\xistar} \pi \left( |\theta_j| \leq a_n \right) \geq (1-p^{-(1+u)})^p \to 1$ as $p \to \infty$ with inequalities above, we have
\[ \pi(\normone{\bm{\theta}-\thetastar}<2\eta\epsilon_n) \geq (2\eta)^{\sstar}p^{-\sstar((k+1)/2+c)} \geq p^{-\sstar(k/2+c+1)}.\]

Thus for $z\in \setEps$ and $c_4 = k/2+c+c_3+1$,
\[
\int_{\Rp}\dfrac{f_{\bm\theta}(z)}{f_{\thetastar}(z)}\pi({\bm\theta})d{\bm\theta} \geq \exp\{-c_4n\epsilon_n^2\}
.\]
We have shown that $\Pstar(z\notin \setEps)\leq 2/p^2$ in \eqref{existtest:1_3}, therefore the lemma is proved. 

\end{proof}

\begin{proof}(Theorem 1)
For this Theorem, we will only consider design matrix $\bX$ that satisfies all the three inequalities in Lemma \ref{lemma:inequalities}, which has probability at least $1-\exp(-c_5 n)$ for some $c_5>0$.
We can verify that $\setBn \supseteq \{{\bm\theta}\in\Rp: \sum_{j=1}^p I(|\theta_j-{\thetastar}_j|>a_n) > \sbar\}$. And note that 
$\setCn = \{\theta\in\Rp: \sum_{j=1}^p I(|\theta_j-{\thetastar}_j|>a_n) \leq \sbar, \normtwo{\theta-\thetastar}>M\epsilon_n\}$, 
therefore 
\[ 
\setBn \cup \setCn = \{{\bm\theta}\in\Rp: \sum_{j=1}^p I(|\theta_j|>a_n) > \sbar-\sstar \text{ or } \normtwo{{\bm\theta}-\thetastar}>M\epsilon_n\}.
\]
By Lemma \ref{lemma:prior_con}, $\pi(\setBn) \leq \exp\{-uL n\epsilon_n^2\}$. Lemma \ref{lemma:existtest} shows existence of test function $\phi$ satisfying
$\E^*[\phi(\bZ)]\leq 4/p^2$ and $\sup_{\theta\in \setCn} \Etheta[1-\phi(\bZ)]\leq  \exp(-c_2n\epsilon_n^2)$. And Lemma \ref{lemma:marginal} shows
\[
\Pstar\left(\int_{\Rp} \dfrac{f_{\bm\theta}(\bZ)}{f_{\thetastar}(\bZ)}\pi(\bm\theta)d\bm\theta \geq \exp(-c_4 n\epsilon_n^2) \right) > 1-\dfrac{2}{p^2}.
\]
Applying Lemma \ref{lemma:song}, we find that
\[
\Estar[\Pi(\setBn\cup\setCn|\bZ)] \leq \dfrac{\exp(-uL n\epsilon_n^2) + \exp(-c_2n\epsilon_n^2)}{\exp(-c_4 n\epsilon_n^2)} + \dfrac{6}{p^2}.
\]
Therefore, $ \Estar[\Pi(\setBn\cup\setCn|Z)] \leq 7/p^2 $ as long as $c_4 \leq \min(uL, c_2)-2$. Since $c_2$ can be very large, the result holds if $L \geq (c_4+2)/u$.
\end{proof}

\begin{proof} (Theorem 2)
Following the notations in Lemma \ref{lemma:inequalities}, let $\setG_p=\setG_{1,p}\cap \setG_{2,p} \cap \setG_{3,p}$, where
\begin{align*}
    \setG_{1,p} & :=\{X: \dfrac{\normtwo{X v}}{\sqrt{n}} \geq \dfrac{1}{4}\normtwo{\Sigma^{1/2}v} - 9\bar{\sigma}\sqrt{\dfrac{\log p}{n}}\normone{v} \text{for all } v\in\Rp \}, \\
    \setG_{2,p} & :=\{X:  \lambda_{\max}(X^T X) \leq (2\sqrt{n} + \tilde{c}\sqrt{p})^2 \}, \\
    \setG_{3,p} & :=\{X: \max_i \normtwo{X_{.i}}^2 \leq 2\bar{\sigma}^2n \} .
\end{align*}
By Lemma \ref{lemma:inequalities}, $\P(\bY \notin \setG_p)\leq c'\exp(-cn) + 2\exp(-\tilde{c}'n) + p \exp(-n/8) \leq \exp(-c_5n)$ for large $n$, where $c_5>0$ is a constant free of $n$ and $p$.

Let $X^{(j)} \in \R^{n\times(p-1)}$ denote the matrix obtained by removing the $j$-th column of $X$.
It is not hard to see that if $X \in \setG_{j,p}$, then $X^{(j)} \in \setG_{j,p-1}, \;j=1,2,3$. 
Therefore, we have
\begin{align*}
     & \EstarOmega[\tilPi(\{\Omega: \sum_{i=1, i\neq j}^p I(\omega_{ij}>a_n)  \geq L\sstar_j \text{ for some } j\} |\bY)]  \\
\leq & \P(\bY \notin \setG_p) + 
     \sum_{j=1}^p\EstarOmega \left[ 
     I(\bY\in\setG_p) \tilPi_j(\{\Omegajcol: \sum_{i=1, i\neq j}^p I(\omega_{ij}>a_n)  \geq L\sstar_j\} |\bY) \right] \\
\leq & \exp(-c_5n) + 
     \sum_{j=1}^p\EstarOmega \left[I(\bY_{\cdot, -j} \in\setG_{p-1}) T_j \right]
\end{align*}
where $T_j = \E\left(\tilPi_j( \{\Omegajcol: \sum_{i=1, i\neq j}^p I(\omega_{ij}>a_j)  \geq L\sstar_j\}|\bY) | \bY_{\cdot, -j} \right)$ and $a_j=\sqrt{\sstar_j\log p/n}/p$.

Note that $T_j$ is the expected quasi-posterior probability \citep{atchade2019quasi} in the following linear regression model 
\[  
\bY_{\cdot,j} = \bY_{\cdot, -j}\Omega_{-j,j} + \bm{\varepsilon}, \; \bm{\varepsilon}\sim N(0,I_n).
\]
By \eqref{pstcon:linear_selection} in Theorem \ref{thm:linearreg}, we have $T_j \leq 7/p^2$ for $\bY_{\cdot, -j}\in\setG_{p-1}$, hence
\[
\EstarOmega[\tilPi(\{\Omega: \sum_{i=1, i\neq j}^p I(\omega_{ij}>a_n) \geq L\sstar_j \text{ for some } j \}|\bY)] 
\leq \exp(-c_5 n) + 7/p. 
\]
Using the same idea, let $U_j = \E\left( \tilPi_j( \{\Omegajcol: \normtwo{\Omega_{.j}-{\Omegastar}_{.j}} > M\epsilon_j\} |\bY) | \bY_{\cdot, -j} \right)$, where $\epsilon_j=\sqrt{\sstar_j\log p/n }$. Then $U_j\leq 7/p^2$ by Theorem \ref{thm:linearreg} \eqref{pstcon:linear_est}. Thus we have
\begin{align*}
      & \EstarOmega[\tilPi(\{\Omega: \normF{\Omega-\Omegastar} > M\epsilon_n\} | \bY)]  \\
 \leq & \EstarOmega[\tilPi(\{\Omega: \normtwo{\Omega_{.j}-{\Omegastar}_{.j}} > M\epsilon_j \text{ for some } j\} | \bY)] \\
 \leq & \P(\bY \notin \setG_p) + \sum_{j=1}^p\EstarOmega \left[I(\bY_{\cdot, -j}\in\setG_{p-1}) U_j \right] \\
 \leq & \exp(-c_5 n) + 7/p. 
\end{align*}
\end{proof}

\begin{proof}(Corollary 2.1)
For the symmetric matrix $\hat{\Omega}-\Omegastar$, we have
\[
\norm{\hat{\Omega}-\Omegastar}_{\ell_2} \leq \norm{\hat{\Omega}-\Omegastar}_{\ell_1} = \norm{\hat{\Omega}-\Omega + \Omega -\Omegastar}_{\ell_1} \leq 2 \norm{\Omega -\Omegastar}_{\ell_1},
\]
where $\Omega$ is a random sample from the quasi-posterior distribution $\tilPi(\cdot|\bY)$.

For $\Omega$'s $j$-th column $\Omegajcol$, define $\gamma_{a_j}(\Omegajcol) = \sum_{i=1, i\neq j}^p I(|\Omega_{ij}|>a_j)$, where $a_j=\epsilon_j/p$ and $\epsilon_j = \sqrt{\sstar_j \log p/n}$. Then we have,
\[
\norm{\Omega -\Omegastar}_{\ell_1} = \max_{j} \normone{\Omegajcol-\Omegastar{_{\cdot , j}}}
\leq  \max_{j} \left(\epsilon_j +  \sqrt{\gamma_{a_j}(\Omegajcol) + \sstar_j}\normtwo{\Omegajcol-\Omegastar{_{\cdot,j}}} \right).
\]
\begin{align*}
& \EstarOmega[\tilPi(\norm{\hat\Omega-\Omegastar}_{\ell_2} > M d^*\sqrt{\log p/n} | \bY)] \\
\leq & \EstarOmega[\tilPi( \max_{j} ( \epsilon_j +  \sqrt{\gamma_{a_j}(\Omegajcol) + \sstar_j}\normtwo{\Omegajcol-\Omegastar{_{\cdot,j}}} ) > \frac{M}{2} d^*\sqrt{\log p/n} | \bY)] \\
\leq & \P(\bY \notin \setG_p) + 
     \sum_{j=1}^p \EstarOmega[\tilPi(\sqrt{\gamma_{a_j}(\Omegajcol) + \sstar_j}\normtwo{\Omegajcol-\Omegastar{_{\cdot,j}}} > (\frac{M}{2}-1) \sstar_j\sqrt{\log p/n} | \bY)
     I(\bY\in\setG_p)] \\
\leq & \exp(-c_5 n) + \sum_{j=1}^p\EstarOmega \left[I(\bY_{\cdot, -j}\in\setG_{p-1}) (T_j + \tilde{U}_j) \right],
\end{align*}
where $T_j$ is defined above in proof of Theorem \ref{thm:ggm} and 
\[
\tilde{U}_j = \E\left( \tilPi_j( \{\Omegajcol: \normtwo{\Omega_{.j}-{\Omegastar}_{.j}} > \frac{M/2-1}{\sqrt{L+1}}\epsilon_j\} |\bY) | \bY_{\cdot, -j} \right).
\]
For $\bY_{\cdot, -j}\in\setG_{p-1}$ and sufficient large constant $M$, $T_j \leq 7/p^2$ and $\tilde{U}_j \leq 7/p^2$. Hence,
\[
\EstarOmega[\tilPi(\norm{\hat\Omega-\Omegastar}_{\ell_2} > M d^*\sqrt{\log p/n} | \bY)] \leq \exp(-c_5 n) + 14/p.
\]
\end{proof}

\end{document}


\section{Supplementary Material}
\subsection{Gibbs Sampler for quasiGHS}
The horseshoe prior imposed on the quasiGHS model is:
\begin{align*}
    \omega_{ij}| \lambda_{ij},\tau & \sim N(0,\lambda_{ij}^2\tau^2), \; i\neq j \\
    \lambda_{ij} &\sim C^+(0,1) \\
    \tau &\sim C^+(0,1).
\end{align*}    

\noindent With scale mixture representation of half-Cauchy distribution\cite{makalic2015simple}, distribution of $\lambda_{ij}$ and $\tau$ can be written as:
\begin{align*}
    \lambda_{ij}^2|v_{ij} &\sim IG(1/2, 1/v_{ij})\\
    \tau^2|\kappa &\sim IG(1/2,1/k)\\
    v_{ij}, \kappa &\sim IG(1/2,1).
\end{align*}    
The data pseudo-likelihood is as follows:
\begin{align*}
     \bY | \Omega \propto & 
     \prod_{i=1}^p \omega_{ii}^{n/2}\exp\{-\dfrac{\omega_{ii}}{2}\normtwo{\bY_i + \sum_{j\neq i}\dfrac{\omega_{ji}}{\omega_{ii}}\bY_j}^2\}\\
     = &  \prod_{i=1}^p \omega_{ii}^{n/2}\exp\{-\dfrac{1}{2\omega_{ii}}\normtwo{\omega_{ii}\bY_i + \sum_{j\neq i}\omega_{ji}\bY_j}^2\}.
\end{align*}
Hence the gibbs sampler for quasiHS is:
\begin{align*}
    \omega_{ji} | else &\sim N(-\frac{\sum\limits_{k\neq j} \frac{\omega_{ki}}{\omega_{ii}}s_{kj}}{\frac{s_{jj}}{\omega_{ii}} + \frac{1}{\tau^2\lambda_{ji}^2}}, \frac{1}{\frac{s_{jj}}{\omega_{ii}} + \frac{1}{\tau^2\lambda_{ji}^2}}), \; j \neq i\\
    \dfrac{1}{\lambda_{ji}^2} | else &\sim Exp(rate=\frac{\omega_{ji}^2}{2\tau^2}+\frac{1}{v_{ji}})\\
    \dfrac{1}{v_{ji}} | else &\sim Exp(rate=1+\frac{1}{\lambda_{ji}^2})\\
    \dfrac{1}{\tau^2} | else &\sim Gamma(shape=\frac{p(p-1)+1}{2}, rate=\frac{1}{k}+\frac{1}{2}\sum\limits_{j \neq i}\frac{\omega_{ji}^2}{\lambda_{ji}^2}) \\
    \dfrac{1}{\kappa} | else &\sim Exp(rate=1+\frac{1}{\tau^2}).
\end{align*}

\subsection{MCMC Convergence Diagnostics}
The trace plot of Frobenius norm across iterations Figure demonstrates the convergence of the four Gibbs sampler of BGL, GHS, quasiGHS-diag and quasiGHS. 
We evaluate the convergence and mixing of the proposed graphical horseshoe , using different starting values for each chain. It is shown that when p = 100, n=150, MCMC samples converge within 500 iterations, and mix reasonably well. We use 1000 burn-in samples. 
We use 5000 iterations in all cases, for both graphical horseshoe and Bayesian graphical lasso. Figure S.1 shows the plots used for MCMC diagnostics. Gelman–Rubin diagnostics test is given in table. 


\bibliographystyle{plain}
\bibliography{Pseudo_GGM.bib}

%% file: mathcommand.tex
\usepackage{amsmath,amssymb,amsfonts,bm}

\DeclareMathOperator*{\argmin}{arg\,min}

\def\sbar{\bar{s}}
\def\sstar{s^*}
\def\thetastar{\bm{\theta}^*}
\def\thetakl{\theta_{k,l}}

\def\ck{\tilde{c}k_1}

\def\xistar{\xi_*}

\def\tr{\mathrm{tr}}
\def\DE{\mathrm{DE}}
\def\EXP{\mathrm{EXP}}
\def\halfCauchy{\mathrm{C^+}}

\def\bv{\bm{v}}

\def\Omegastar{\Omega_*}

\def\Omegajcol{\Omega_{\cdot, j}}

\def\R{\mathbb{R}}
\def\Rp{\mathbb{R}^p}
\def\Rnp{\mathbb{R}^{n\times p}}
\def\splspcZn{\mathcal{Z}^{(n)}}

\def\tilPi{\tilde{\Pi}}

\def\bigO{\mathcal{O}}

\def\pialpha{\pi_\alpha}

\def\bX{\bm{X}}
\def\bx{\bm{x}}

\def\bY{\bm{Y}}
\def\bZ{\bm{Z}}

\def\fstar{f_{\theta_*}}
\def\htheta{h_\theta}

\def\setA{\mathcal{A}}
\def\setTilTheta{\tilde{\Theta}}
\def\setTilEps{\tilde{\mathcal{E}}}
\def\setEps{\mathcal{E}}
\def\setBallp{\mathcal{B}_p}
\def\setSl{\mathcal{S}_l}
\def\setQthetak{\mathcal{Q}_{\thetakl}}
\def\setConvQ{conv(\setQthetak)}
\def\setCn{\mathcal{C}_n}
\def\setBn{\mathcal{B}_n}
\def\setG{\mathcal{G}}
\def\setS{\mathcal{S}}

\def\P{\mathbb{P}}
\def\Pstar{\mathbb{P}^*}

\def\E{\mathbb{E}}
\def\Estar{\mathbb{E}_{\theta_*}}
\def\Etheta{\mathbb{E}_\theta}
\def\EstarOmega{\mathbb{E}_{\Omega_*}}
\def\Hhalf{\mathcal{H}_{1/2}}

\DeclarePairedDelimiter{\norm}{\lVert}{\rVert}
\DeclarePairedDelimiter{\normzero}{\lVert}{\rVert_0}
\DeclarePairedDelimiter{\normone}{\lVert}{\rVert_1}
\DeclarePairedDelimiter{\normtwo}{\lVert}{\rVert_2}
\DeclarePairedDelimiter{\norminf}{\lVert}{\rVert_\infty}
\DeclarePairedDelimiter{\normF}{\lVert}{\rVert_F}

%% file: main.bbl
\begin{thebibliography}{}

\bibitem[Armagan et~al., 2011]{armagan2011generalized}
Armagan, A., Clyde, M., and Dunson, D. (2011).
\newblock Generalized beta mixtures of gaussians.
\newblock In {\em Advances in Neural Information Processing Systems},
  volume~24.

\bibitem[Armagan et~al., 2013a]{armagan2013generalized}
Armagan, A., Dunson, D.~B., and Lee, J. (2013a).
\newblock Generalized double pareto shrinkage.
\newblock {\em Statistica Sinica}, 23(1):119.

\bibitem[Armagan et~al., 2013b]{armagan2013posterior}
Armagan, A., Dunson, D.~B., Lee, J., Bajwa, W.~U., and Strawn, N. (2013b).
\newblock Posterior consistency in linear models under shrinkage priors.
\newblock {\em Biometrika}, 100(4):1011--1018.

\bibitem[Atchad{\'e}, 2017]{atchade2017contraction}
Atchad{\'e}, Y.~A. (2017).
\newblock On the contraction properties of some high-dimensional
  quasi-posterior distributions.
\newblock {\em Annals of Statistics}, 45(5):2248--2273.

\bibitem[Atchad{\'e}, 2019]{atchade2019quasi}
Atchad{\'e}, Y.~F. (2019).
\newblock Quasi-bayesian estimation of large gaussian graphical models.
\newblock {\em Journal of Multivariate Analysis}, 173:656--671.

\bibitem[Banerjee et~al., 2008]{banerjee2008model}
Banerjee, O., El~Ghaoui, L., and d'Aspremont, A. (2008).
\newblock Model selection through sparse maximum likelihood estimation for
  multivariate gaussian or binary data.
\newblock {\em The Journal of Machine Learning Research}, 9:485--516.

\bibitem[Banerjee and Ghosal, 2014]{banerjee2014posterior}
Banerjee, S. and Ghosal, S. (2014).
\newblock Posterior convergence rates for estimating large precision matrices
  using graphical models.
\newblock {\em Electronic Journal of Statistics}, 8(2):2111--2137.

\bibitem[Banerjee and Ghosal, 2015]{banerjee2015bayesian}
Banerjee, S. and Ghosal, S. (2015).
\newblock Bayesian structure learning in graphical models.
\newblock {\em Journal of Multivariate Analysis}, 136:147--162.

\bibitem[Belitser and Ghosal, 2020]{belitser2020empirical}
Belitser, E. and Ghosal, S. (2020).
\newblock Empirical bayes oracle uncertainty quantification for regression.
\newblock {\em Annals of Statistics}, 48(6):3113--3137.

\bibitem[Bhadra et~al., 2017]{bhadra2017horseshoe+}
Bhadra, A., Datta, J., Polson, N.~G., and Willard, B. (2017).
\newblock The horseshoe+ estimator of ultra-sparse signals.
\newblock {\em Bayesian Analysis}, 12(4):1105--1131.

\bibitem[Bhadra and Mallick, 2013]{bhadra2013joint}
Bhadra, A. and Mallick, B.~K. (2013).
\newblock Joint high-dimensional bayesian variable and covariance selection
  with an application to eqtl analysis.
\newblock {\em Biometrics}, 69(2):447--457.

\bibitem[Bhattacharya et~al., 2015]{bhattacharya2015dirichlet}
Bhattacharya, A., Pati, D., Pillai, N.~S., and Dunson, D.~B. (2015).
\newblock Dirichlet--laplace priors for optimal shrinkage.
\newblock {\em Journal of the American Statistical Association},
  110(512):1479--1490.

\bibitem[Breiman, 1996]{breiman1996heuristics}
Breiman, L. (1996).
\newblock Heuristics of instability and stabilization in model selection.
\newblock {\em Annals of Statistics}, 24(6):2350--2383.

\bibitem[Brown and Griffin, 2010]{brown2010inference}
Brown, P.~J. and Griffin, J.~E. (2010).
\newblock Inference with normal-gamma prior distributions in regression
  problems.
\newblock {\em Bayesian analysis}, 5(1):171--188.

\bibitem[Candes and Tao, 2007]{candes2007dantzig}
Candes, E. and Tao, T. (2007).
\newblock The dantzig selector: Statistical estimation when p is much larger
  than n.
\newblock {\em Annals of statistics}, 35(6):2313--2351.

\bibitem[Carvalho et~al., 2010]{carvalho2010horseshoe}
Carvalho, C.~M., Polson, N.~G., and Scott, J.~G. (2010).
\newblock The horseshoe estimator for sparse signals.
\newblock {\em Biometrika}, 97(2):465--480.

\bibitem[Carvalho and Scott, 2009]{carvalho2009objective}
Carvalho, C.~M. and Scott, J.~G. (2009).
\newblock Objective bayesian model selection in gaussian graphical models.
\newblock {\em Biometrika}, 96(3):497--512.

\bibitem[Castillo et~al., 2015]{castillo2015bayesian}
Castillo, I., Schmidt-Hieber, J., and Van~der Vaart, A. (2015).
\newblock Bayesian linear regression with sparse priors.
\newblock {\em Annals of Statistics}, 43(5):1986--2018.

\bibitem[Castillo and van~der Vaart, 2012]{castillo2012needles}
Castillo, I. and van~der Vaart, A. (2012).
\newblock Needles and straw in a haystack: Posterior concentration for possibly
  sparse sequences.
\newblock {\em The Annals of Statistics}, 40(4):2069--2101.

\bibitem[Dempster, 1972]{dempster1972covariance}
Dempster, A.~P. (1972).
\newblock Covariance selection.
\newblock {\em Biometrics}, pages 157--175.

\bibitem[Edwards, 2012]{edwards2012introduction}
Edwards, D. (2012).
\newblock {\em Introduction to graphical modelling}.
\newblock Springer Science \& Business Media.

\bibitem[Fan et~al., 2009]{fan2009network}
Fan, J., Feng, Y., and Wu, Y. (2009).
\newblock Network exploration via the adaptive lasso and scad penalties.
\newblock {\em The annals of applied statistics}, 3(2):521.

\bibitem[Fan and Li, 2001]{fan2001variable}
Fan, J. and Li, R. (2001).
\newblock Variable selection via nonconcave penalized likelihood and its oracle
  properties.
\newblock {\em Journal of the American statistical Association},
  96(456):1348--1360.

\bibitem[Fan et~al., 2016]{fan2016overview}
Fan, J., Liao, Y., and Liu, H. (2016).
\newblock An overview of the estimation of large covariance and precision
  matrices.
\newblock {\em The Econometrics Journal}, 19(1):C1--C32.

\bibitem[Friedman et~al., 2008]{friedman2008sparse}
Friedman, J., Hastie, T., and Tibshirani, R. (2008).
\newblock Sparse inverse covariance estimation with the graphical lasso.
\newblock {\em Biostatistics}, 9(3):432--441.

\bibitem[Friedman et~al., 2010]{friedman2010applications}
Friedman, J., Hastie, T., and Tibshirani, R. (2010).
\newblock Applications of the lasso and grouped lasso to the estimation of
  sparse graphical models.
\newblock Technical report, Technical report, Stanford University.

\bibitem[Galloway, 2018]{rpackage_cvglasso}
Galloway, M. (2018).
\newblock {\em CVglasso: Lasso Penalized Precision Matrix Estimation}.
\newblock R package version 1.0.

\bibitem[Ghosal, 1997]{ghosal1997normal}
Ghosal, S. (1997).
\newblock Normal approximation to the posterior distribution for generalized
  linear models with many covariates.
\newblock {\em Mathematical Methods of Statistics}, 6(3):332--348.

\bibitem[Ghosal, 1999]{ghosal1999asymptotic}
Ghosal, S. (1999).
\newblock Asymptotic normality of posterior distributions in high-dimensional
  linear models.
\newblock {\em Bernoulli}, 5(2):315--331.

\bibitem[Ghosal, 2000]{ghosal2000asymptotic}
Ghosal, S. (2000).
\newblock Asymptotic normality of posterior distributions for exponential
  families when the number of parameters tends to infinity.
\newblock {\em Journal of Multivariate Analysis}, 74(1):49--68.

\bibitem[Ghosal et~al., 2000]{ghosal2000convergence}
Ghosal, S., Ghosh, J.~K., and Van Der~Vaart, A.~W. (2000).
\newblock Convergence rates of posterior distributions.
\newblock {\em Annals of Statistics}, pages 500--531.

\bibitem[Hans, 2009]{hans2009bayesian}
Hans, C. (2009).
\newblock Bayesian lasso regression.
\newblock {\em Biometrika}, 96(4):835--845.

\bibitem[Huang et~al., 2006]{huang2006covariance}
Huang, J.~Z., Liu, N., Pourahmadi, M., and Liu, L. (2006).
\newblock Covariance matrix selection and estimation via penalised normal
  likelihood.
\newblock {\em Biometrika}, 93(1):85--98.

\bibitem[Jiang, 2007]{jiang2007bayesian}
Jiang, W. (2007).
\newblock Bayesian variable selection for high dimensional generalized linear
  models: convergence rates of the fitted densities.
\newblock {\em The Annals of Statistics}, 35(4):1487--1511.

\bibitem[Khare et~al., 2015]{khare2015convex}
Khare, K., Oh, S.-Y., and Rajaratnam, B. (2015).
\newblock A convex pseudolikelihood framework for high dimensional partial
  correlation estimation with convergence guarantees.
\newblock {\em Journal of the Royal Statistical Society: Series B (Statistical
  Methodology)}, 77(4):803--825.

\bibitem[Lam and Fan, 2009]{lam2009sparsistency}
Lam, C. and Fan, J. (2009).
\newblock Sparsistency and rates of convergence in large covariance matrix
  estimation.
\newblock {\em Annals of statistics}, 37(6B):4254.

\bibitem[Lange et~al., 2000]{lange2000optimization}
Lange, K., Hunter, D.~R., and Yang, I. (2000).
\newblock Optimization transfer using surrogate objective functions.
\newblock {\em Journal of computational and graphical statistics}, 9(1):1--20.

\bibitem[Lauritzen, 1996]{lauritzen1996graphical}
Lauritzen, S.~L. (1996).
\newblock {\em Graphical models}, volume~17.
\newblock Clarendon Press.

\bibitem[Lee and Lee, 2021]{lee2021estimating}
Lee, K. and Lee, J. (2021).
\newblock Estimating large precision matrices via modified cholesky
  decomposition.
\newblock {\em Statistica Sinica}, 31(2021):173--196.

\bibitem[Li et~al., 2019]{li2019graphical}
Li, Y., Craig, B.~A., and Bhadra, A. (2019).
\newblock The graphical horseshoe estimator for inverse covariance matrices.
\newblock {\em Journal of Computational and Graphical Statistics},
  28(3):747--757.

\bibitem[Liu and Wang, 2017]{liu2017tiger}
Liu, H. and Wang, L. (2017).
\newblock Tiger: A tuning-insensitive approach for optimally estimating
  gaussian graphical models.
\newblock {\em Electronic Journal of Statistics}, 11(1):241--294.

\bibitem[Meinshausen and B{\"u}hlmann, 2006]{meinshausen2006high}
Meinshausen, N. and B{\"u}hlmann, P. (2006).
\newblock High-dimensional graphs and variable selection with the lasso.
\newblock {\em Annals of statistics}, 34(3):1436--1462.

\bibitem[Park and Casella, 2008]{park2008bayesian}
Park, T. and Casella, G. (2008).
\newblock The bayesian lasso.
\newblock {\em Journal of the American Statistical Association},
  103(482):681--686.

\bibitem[Polson and Scott, 2010]{polson2010shrink}
Polson, N.~G. and Scott, J.~G. (2010).
\newblock Shrink globally, act locally: Sparse bayesian regularization and
  prediction.
\newblock {\em Bayesian statistics}, 9(501-538):105.

\bibitem[Polson and Scott, 2012]{polson2012half}
Polson, N.~G. and Scott, J.~G. (2012).
\newblock On the half-cauchy prior for a global scale parameter.
\newblock {\em Bayesian Analysis}, 7(4):887--902.

\bibitem[Raskutti et~al., 2010]{raskutti2010restricted}
Raskutti, G., Wainwright, M.~J., and Yu, B. (2010).
\newblock Restricted eigenvalue properties for correlated gaussian designs.
\newblock {\em The Journal of Machine Learning Research}, 11:2241--2259.

\bibitem[Ravikumar et~al., 2011]{ravikumar2011high}
Ravikumar, P., Wainwright, M.~J., Raskutti, G., and Yu, B. (2011).
\newblock High-dimensional covariance estimation by minimizing
  $\ell_1$-penalized log-determinant divergence.
\newblock {\em Electronic Journal of Statistics}, 5:935--980.

\bibitem[Rocha et~al., 2008]{rocha2008path}
Rocha, G.~V., Zhao, P., and Yu, B. (2008).
\newblock A path following algorithm for sparse pseudo-likelihood inverse
  covariance estimation (splice).
\newblock {\em arXiv preprint arXiv:0807.3734}.

\bibitem[Ro{\v{c}}kov{\'a}, 2018]{rovckova2018bayesian}
Ro{\v{c}}kov{\'a}, V. (2018).
\newblock Bayesian estimation of sparse signals with a continuous
  spike-and-slab prior.
\newblock {\em The Annals of Statistics}, 46(1):401--437.

\bibitem[Rothman et~al., 2008]{rothman2008sparse}
Rothman, A.~J., Bickel, P.~J., Levina, E., and Zhu, J. (2008).
\newblock Sparse permutation invariant covariance estimation.
\newblock {\em Electronic Journal of Statistics}, 2:494--515.

\bibitem[Shen and Ghosal, 2016]{shen2016adaptive}
Shen, W. and Ghosal, S. (2016).
\newblock Adaptive bayesian density regression for high-dimensional data.
\newblock {\em Bernoulli}, 22(1):396--420.

\bibitem[Song, 2020]{song2020bayesian}
Song, Q. (2020).
\newblock Bayesian shrinkage towards sharp minimaxity.
\newblock {\em Electronic Journal of Statistics}, 14(2):2714--2741.

\bibitem[Song and Liang, 2017]{song2017nearly}
Song, Q. and Liang, F. (2017).
\newblock Nearly optimal bayesian shrinkage for high dimensional regression.
\newblock {\em arXiv preprint arXiv:1712.08964}.

\bibitem[Sun and Zhang, 2012]{sun2012scaled}
Sun, T. and Zhang, C.-H. (2012).
\newblock Scaled sparse linear regression.
\newblock {\em Biometrika}, 99(4):879--898.

\bibitem[Sun and Zhang, 2013]{sun2013sparse}
Sun, T. and Zhang, C.-H. (2013).
\newblock Sparse matrix inversion with scaled lasso.
\newblock {\em The Journal of Machine Learning Research}, 14(1):3385--3418.

\bibitem[Terenin et~al., 2019]{terenin2019gpu}
Terenin, A., Dong, S., and Draper, D. (2019).
\newblock Gpu-accelerated gibbs sampling: a case study of the horseshoe probit
  model.
\newblock {\em Statistics and Computing}, 29(2):301--310.

\bibitem[Tibshirani, 1996]{tibshirani1996regression}
Tibshirani, R. (1996).
\newblock Regression shrinkage and selection via the lasso.
\newblock {\em Journal of the Royal Statistical Society: Series B
  (Methodological)}, 58(1):267--288.

\bibitem[Van Der~Pas et~al., 2014]{van2014horseshoe}
Van Der~Pas, S.~L., Kleijn, B.~J., and Van Der~Vaart, A.~W. (2014).
\newblock The horseshoe estimator: Posterior concentration around nearly black
  vectors.
\newblock {\em Electronic Journal of Statistics}, 8(2):2585--2618.

\bibitem[Vershynin, 2010]{vershynin2010introduction}
Vershynin, R. (2010).
\newblock Introduction to the non-asymptotic analysis of random matrices.
\newblock {\em arXiv preprint arXiv:1011.3027}.

\bibitem[Wang, 2012]{wang2012bayesian}
Wang, H. (2012).
\newblock Bayesian graphical lasso models and efficient posterior computation.
\newblock {\em Bayesian Analysis}, 7(4):867--886.

\bibitem[Wei and Ghosal, 2020]{wei2020contraction}
Wei, R. and Ghosal, S. (2020).
\newblock Contraction properties of shrinkage priors in logistic regression.
\newblock {\em Journal of Statistical Planning and Inference}, 207:215--229.

\bibitem[Whittaker, 1990]{whittaker1990graphical}
Whittaker, J. (1990).
\newblock {\em Graphical models in applied multivariate statistics}, volume~19.
\newblock Wiley Chichester.

\bibitem[Williams, 2020]{rpackage_ggmncv}
Williams, D. (2020).
\newblock {\em GGMncv: Gaussian Graphical Models with Nonconvex
  Regularization}.
\newblock R package version 2.0.0.

\bibitem[Xiang et~al., 2015]{xiang2015high}
Xiang, R., Khare, K., and Ghosh, M. (2015).
\newblock High dimensional posterior convergence rates for decomposable
  graphical models.
\newblock {\em Electronic Journal of Statistics}, 9(2):2828--2854.

\bibitem[Yang and Tokdar, 2015]{yang2015minimax}
Yang, Y. and Tokdar, S.~T. (2015).
\newblock Minimax-optimal nonparametric regression in high dimensions.
\newblock {\em Annals of Statistics}, 43(2):652--674.

\bibitem[Yuan, 2010]{yuan2010high}
Yuan, M. (2010).
\newblock High dimensional inverse covariance matrix estimation via linear
  programming.
\newblock {\em The Journal of Machine Learning Research}, 11:2261--2286.

\bibitem[Yuan and Lin, 2007]{yuan2007model}
Yuan, M. and Lin, Y. (2007).
\newblock Model selection and estimation in the gaussian graphical model.
\newblock {\em Biometrika}, 94(1):19--35.

\bibitem[Zhang and Ghosh, 2019]{zhang2019ultra}
Zhang, R. and Ghosh, M. (2019).
\newblock Ultra high-dimensional multivariate posterior contraction rate under
  shrinkage priors.
\newblock {\em arXiv preprint arXiv:1904.04417}.

\bibitem[Zhang and Zou, 2014]{zhang2014sparse}
Zhang, T. and Zou, H. (2014).
\newblock Sparse precision matrix estimation via lasso penalized d-trace loss.
\newblock {\em Biometrika}, 101(1):103--120.

\bibitem[Zhao and Liu, 2014]{zhao2014calibrated}
Zhao, T. and Liu, H. (2014).
\newblock Calibrated precision matrix estimation for high-dimensional
  elliptical distributions.
\newblock {\em IEEE transactions on Information Theory}, 60(12):7874--7887.

\end{thebibliography}
